%% file: arxiv.tex
\documentclass[journal]{IEEEtran}
\usepackage{comment}
\usepackage{cite} 

\ifCLASSINFOpdf 
\else
\fi

\usepackage{amsmath} 
\usepackage{algorithmic} 
\usepackage{array} 
\usepackage[table,dvipsnames]{xcolor}
\usepackage{dblfloatfix}
\usepackage{url} 
\usepackage{nomencl}
\usepackage[acronym,nonumberlist,toc,nogroupskip]{glossaries}
\usepackage{tkz-euclide}
\usepackage{import}
\usepackage{color}
\usepackage{xcolor}
\usepackage{lipsum}
\usepackage{hhline}
\usepackage{multirow}
\usepackage{longtable}
\usepackage{amsmath}
\usepackage{tabularx}
\usepackage{amsthm,latexsym,amssymb,amsmath, amsfonts}
\usepackage{graphicx}
\usepackage{array,amsmath}
\usepackage{longtable}
\usepackage{multirow}
\usepackage[justification=centering,singlelinecheck=false,font=footnotesize,skip=4pt]{caption}
\usepackage[justification=centering,font=footnotesize]{subcaption}
\setglossarystyle{super}

\newacronym{ess}{ESS}{Energy Storage System}
\newacronym{fr}{FR}{Frequency Regulation}
\newacronym{bess}{BESS}{Battery Energy Storage System}
\newacronym{fess}{FESS}{Flywheel Energy Storage System}
\newacronym{SoC}{SoC}{State of Charge}
\newacronym{dsatools}{DSATools\texttrademark}{Dynamic Security Assessment Tools}
\newacronym{ace}{ACE}{Area Control Error}
\newacronym{iso}{ISO}{Independent System Operator}
\newacronym{rto}{RTO}{Regional Transmission Organization}
\newacronym{ferc}{FERC}{Federal Energy Regulatory Commission}
\newacronym{cps1}{CPS1}{Control Performance Standard 1}
\newacronym{ieso}{IESO}{Independent Electricity System Operator}
\newacronym{pjm}{PJM}{PJM Interconnection LLC}
\newacronym{scada}{SCADA}{Supervisory Control and Data Acquisition System}
\newacronym{agc}{AGC}{Automatic Generation Control}
\newacronym{nn}{NN}{Neural Network}
\newacronym{lstm}{LSTM}{Long-Short Term Memory}
\newacronym{tg}{TG}{Traditional Generator}
\newacronym{sp}{SP}{Set-Point}
\newacronym[longplural={Balancing Authorities}]{ba}{BA}{Balancing Authority}
\newacronym{naei}{NAEI}{North American Eastern Interconnection}
\newacronym{rmspe}{RMSPE}{Root-Mean Squared Percentage Error}
\newacronym{mspe}{MSPE}{Mean Squared Percentage Error}
\newacronym{mape}{MAPE}{Mean Absolute Percentage Error}
\newacronym{rmse}{RMSE}{Root-Mean Squared Error}
\newacronym{mse}{MSE}{Root-Mean Squared Error}
\newacronym{mae}{MAE}{Mean Absolute Error}
\newacronym{tsat}{TSAT}{Transient Security Assessment Tool}
\newacronym{gp}{GP}{Generalized Plant}
\newacronym{baal}{BAAL}{Balancing Authority ACE Limit}
\newacronym{atr}{ATR}{Alternate Technologies for Regulation}
\newacronym{rfp}{RFP}{Request for Proposals}
\newacronym{ops}{OPS}{Ontario Power System}
\newacronym{res}{RES}{Renewable Energy Source}
\usepackage{setspace}
\begin{document} 
\markboth{Journal of \LaTeX\ Class Files,~Vol.~14, No.~8, August~2015}%
{Shell \MakeLowercase{\textit{et al.}}: Bare Demo of IEEEtran.cls for IEEE Journals}
\title{Frequency Regulation Model of Bulk Power Systems with Energy Storage}
\author{N. Sofia~Guzman E.,~\IEEEmembership{Member,~IEEE,}
       Claudio A.~Ca{ñ}izares,~\IEEEmembership{Fellow,~IEEE,}
       Kankar Bhattacharya,~\IEEEmembership{Fellow,~IEEE,}
       and~Daniel Sohm,~~\IEEEmembership{Member,~IEEE}
\thanks{This work has been submitted to the IEEE for possible publication. Copyright may be transferred without notice, after which this version may no longer be accessible. This work was supported by the NSERC Energy Storage Technology (NEST) Network.  The energy storage data was kindly provided by NRStor Inc.}
\thanks{N. S. Guzman, C. A. Ca\~nizares, and K. Bhattacharya are with the Electrical \& Computer Engineering Dept., University of Waterloo, Waterloo, ON, N2L 3G1, Canada (e-mail: nguzman@uwaterloo.ca; ccanizares@uwaterloo.ca; kankar.bhattacharya@uwaterloo.ca). D. Sohm is with the IESO of Ontario, Toronto, ON, M5H 1T1, Canada (e-mail: Daniel.Sohm@ieso.ca).}}

\markboth{IEEE Transactions on Power Systems, Submitted September 2020}
{}
\maketitle
\begin{abstract}
This paper presents a dynamic \gls{fr} model of a large interconnected power system including \glspl{ess} such as \glspl{bess} and \glspl{fess}, considering all relevant stages in the frequency control process. Communication delays are considered in the transmission of the signals in the \gls{fr} control loop and \glspl{ess}, and their \gls{SoC} management model is considered. The system, \glspl{ess} and \gls{SoC} components are modelled in detail from a \gls{fr} perspective. The model is validated using real system and \glspl{ess} data, based on a practical transient stability model of the \gls{naei}, and the results show that the proposed model accurately represents the \gls{fr} process of a large interconnected power network including \gls{ess}, and can be used for long-term \gls{fr} studies. The impact of communication delays and \gls{SoC} management of \gls{ess} facilities in the \gls{ace} is also studied and discussed. 
\end{abstract}
\glsresetall

\IEEEpeerreviewmaketitle
\section{Introduction} \label{sec:I}
\IEEEPARstart{H}{IGH} levels of penetration of \glspl{res} are increasing the operational challenges in power grids. Most of the challenges are linked to the uncertainty associated with \gls{res}, which leads to increased generation/load mismatches that particularly impact \gls{fr} and stability. \glspl{ess} can help to maintain grid stability and reliability \cite{8864014,1300713}, providing energy arbitrage, and ancillary services such as \gls{fr}, among others, while being competitive and economically viable \cite{8827655,ES_handbook}.

In recent years, there has been a significant interest in \glspl{ess} because of their decreasing costs \cite{70_cost}. Policies around the world are being modified, and new services are being created to facilitate participation of \glspl{ess} in the electricity markets \cite{8864014,FERC_841,IESO_Es_obstacles}. In some instances, \glspl{iso} are implementing grid-scale \gls{ess} projects to gain experience and evaluate the benefits and performance of different \gls{ess} technologies. For instance, the \gls{atr} program by the \gls{ieso} of Ontario, which includes a \gls{bess} and a \gls{fess} for \gls{fr}, was implemented to provide learning opportunities for all stakeholders \cite{IESO_Es_obstacles}.

The fast power response characteristic of \glspl{ess}, such as \gls{fess} and \gls{bess}, make them particularly suitable to provide fast \gls{fr} services \cite{ES_handbook}. Studies focusing on \glspl{ess} for the provision of \gls{fr} examine their contribution in helping conventional generators \cite{7542188} and \gls{res} \cite{7797224} to meet \gls{fr} requirements, or provide \gls{fr} themselves (e.g. \cite{8359089}). However, appropriate \gls{ess} models for frequency stability studies are lacking, which hampers the impact analysis studies of \glspl{ess} for \gls{fr} from the \gls{iso} perspective. 

The model of a two-area system with \glspl{ess} used for \gls{fr} is presented in \cite{8742681} from the system viewpoint; however, communication delays and \gls{SoC} models of the \gls{ess} facilities are not considered, which may lead to unrealistic effects in the \gls{ace}. The study in \cite{4749312} includes communication delays in the \gls{fr} process, but \glspl{ess} are not included, and while \cite{6719579} considers the \gls{SoC} model of \gls{ess}, the work does not consider communication delays.

Determining the actual and potential benefits of \glspl{ess} for \gls{fr} in an interconnected power system requires a model that represents the overall frequency dynamics of the system and its limitations. The current paper contributes to the on-going efforts of modelling a bulk power system including \gls{ess} technologies for \gls{fr}, as follows:
\begin{itemize}
	\item Develop the \gls{fr} model of a large interconnected power grid, estimating the system model parameters, and validating it with real system information. 
	\item Develop empirically-based \gls{SoC} models for \gls{fess} and \gls{bess} considering the charging and discharging process characteristics, validating them using data from actual facilities.
	\item Analyze the impact of communication delays on the \gls{ace} and the provision of \gls{fr} services by \gls{ess} facilities. 
\end{itemize} 

The rest of the paper is organized as follows: Section~\ref{sec:II} provides a practical overview of frequency control and regulation in power systems, and reviews the \gls{ess} technologies used for such services. Section~\ref{sec:III} presents the proposed system and \gls{ess} \gls{fr} models, and Section \ref{sec:IV} validates the models using real data from the \gls{ops}, two \gls{ess} facilities, and a model of the \gls{naei}. Studies of the impact of communication delays, and the \gls{SoC} management model of \gls{ess} facilities are also presented in Section \ref{sec:IV}. Finally, Section~\ref{sec:conclusions} highlights the main conclusions and contributions of this paper.
\section{Background Review} \label{sec:II}
\subsection{Overview of Frequency Control in Power Systems}\label{subsec:IIA}
Frequency is a system-wide characteristic of power grids and should be maintained within specified limits to ensure the stable and reliable operation of the system. Therefore, appropriate frequency control is essential to maintain the normal operation of the grid. \gls{fr} services are required in order to compensate for forecast errors, non-linear behaviour of demand between dispatches, and generation/load resources that do not follow dispatch instructions. These uncertainties may exceed the contracted capacity of \gls{fr} resources, which is automatically compensated by the grid inter-ties, thereby deviating the interchanges from their scheduled values \cite{NERC_ACE}. 

Primary frequency control is critical for maintaining the reliability of the interconnection after a disturbance, by restoring the generation-load balance, and it is implemented through governor control and automatically assisted by the response of frequency dependent loads \cite{NERC_ACE}. Primary control stabilizes the system frequency at a different value from the scheduled one, with the turbine-governor closed loop control being the main component \cite{Prof_book}. Thus, there is a need to correct the generation-demand mismatch created by steady-state frequency deviations, which is provided by the dispatched generators in the control area and units that could be started up in short time periods. This action, known as secondary control, has the objective to bring the frequency error back to zero to restore the primary control capability of the system, modifying the power reference of the generators that participate in this control, which is referred as \gls{agc}, and maintaining the \gls{ace} within an acceptable range by controlling multiple generators. The \gls{ace} extracts contributions of a control area to the interconnection frequency deviation, monitoring and keeping it within limits \cite{NERC_ACE,Prof_book}.
\subsection{Overview of Frequency Control in Ontario}\label{subsec:IIB}
In order to maintain the reliability of the power system, the \gls{ieso} procures ancillary services, \gls{fr} being one of them. This service seeks to match the generation and load, including losses, to reduce the deviations in the system frequency. Seven generation facilities (hydro and gas) plus two \gls{atr} units are contracted by the \gls{ieso} to provide \gls{fr}. A minimum requirement of $\pm$100 MW of \gls{agc} with a ramp rate of 50 MW/min must be scheduled every hour \cite{Market_rule}. 

Historically, \gls{fr} service has been provided by traditional facilities with \gls{agc}, which change their output in response to the regulation signals. In 2012, the \gls{ieso} included the participation of alternative technologies, such as aggregate loads, \gls{fess}, and \gls{bess} to evaluate their ability to provide \gls{fr} as compared to the existing facilities \cite{IESO_2018}. The \gls{ieso} procured 6 MW for \gls{fr} from two \gls{ess} facilities in 2012, which provide \gls{fr} services exclusively, thus receiving a regulation signal to reduce the system generation-load mismatch in seconds. The reason for including these \gls{atr} units was the need for increasing the \gls{fr} capacity \cite{IESO_2018}, which are being complemented by another 50 MW \gls{ess} capacity under the 2014 Grid Energy Storage Procurement plan \cite{IESO_Es_obstacles}.

In 2015, the \gls{ieso} scheduled a minimum of $\pm$100 MW every hour for \gls{fr} services, a need which was anticipated to increase due to the increasing penetration of intermittent generation, and only compensated for 53\% of the forecast errors; this is expected to further decrease to 40\% by 2020, resulting in increased dependency on the power from tie-lines \cite{IESO_Operability}. In 2016, the \gls{ieso} determined a need for additional \gls{fr} capacity and one of the ways to reach this target is through the 2017 \gls{rfp} for incremental regulation capacity, which seeked to increase the \gls{fr} capacity while being open to different technologies. However, the \gls{ieso} no longer has contracts with the two facilities. A procurement to acquire \gls{fr} capacity is being considered by the \gls{ieso} \cite{IESO_RFP}. 
\subsection{Overview of Energy Storage Systems}\label{subsec:IIC}
\glspl{ess} draw energy from the power system, convert it to another form of energy, and release it back to the system when needed. \glspl{ess} suitable for \gls{fr} are those with fast response characteristic, since they can accurately follow the \gls{fr} signal. Since, \glspl{bess} are able to modify their output in less than one second, it makes them a good fit for \gls{fr} provision, with efficiencies depending on the type (flow or solid state), in the order of 75\% to 95\%; they also have long lifetimes and low energy density \cite{ES_handbook}. Another type of \gls{ess} apt for \gls{fr} provision is \gls{fess}, which is capable of responding in milliseconds to minutes, have long life cycles, insensitivity to deep discharges, and high round-trip efficiency (90-95\%) \cite{ES_handbook}.
\section{Proposed \gls{fr} Model}\label{sec:III}
To determine a baseline \gls{fr} model of a real large interconnected power system and estimate the model parameters for a physical system, for which operational information is available, power system analysis tools, such as \gls{dsatools} used by the \gls{ieso} \cite{dsatools}, can be used, as these provide accurate results by properly modelling the system components. Large number of events in one contingency definition (i.e., load changes and \gls{fr} signal sent to \glspl{tg} and \gls{ess} facilities) with one second between events limits the simulation time; for example, using \gls{dsatools} to model and simulate load changes in the entire \gls{naei} and the \gls{fr} signal sent to a group of generators in one area, limits the maximum simulation time to 102 seconds. However, longer simulation time periods are needed for some frequency studies, such as to calculate the \gls{cps1} or \gls{baal} \cite{baalcps1}. Therefore, the model proposed in this paper and depicted in Fig.~\ref{fig:current_model}, which includes the main stages in the \gls{fr} control process, can be used for accurate long-term studies, i.e., hours, days, or even years. The model presented here was developed with the help of the \gls{ieso}, based on their recommendations and observations of the various signals provided, and through a trial and error approach, using real data. This was implemented in Simulink\textsuperscript{\textregistered} \cite{simulink}, with the parameters within each block being determined using the Parameter Estimator application available in this software.
\subsection{Bulk Power System }\label{subsec:IIIA}
 \begin{figure}[t!]
    \centering
    \def\svgwidth{\columnwidth}
    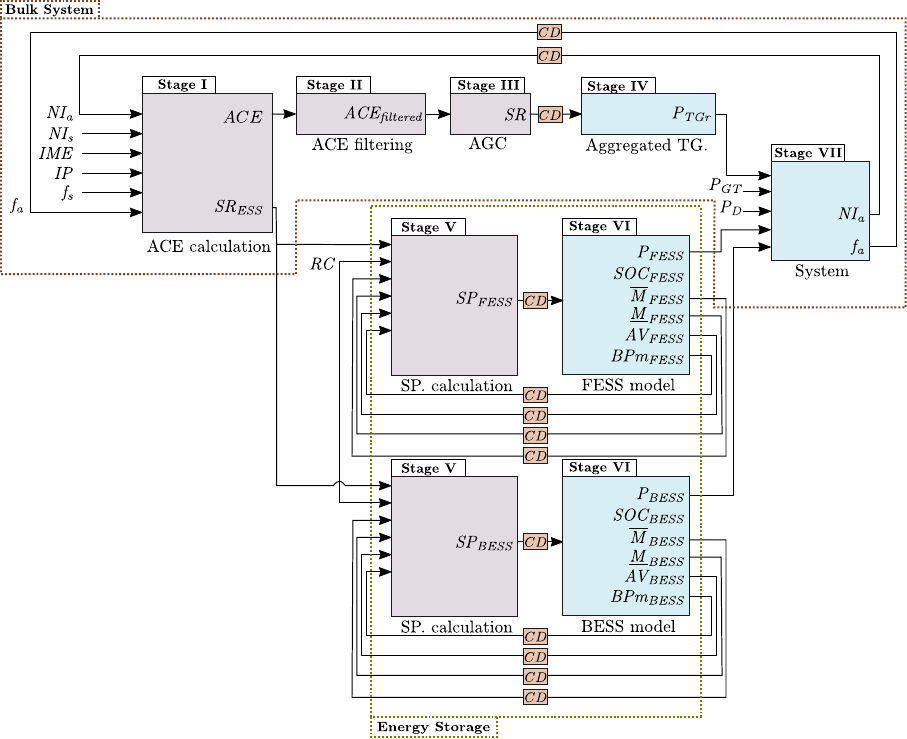
    \caption{Frequency response model of a large interconnected power system with FESS and BESS.}
    \label{fig:current_model}
\end{figure}
The model in Fig.~\ref{fig:current_model} is based on \gls{ieso}'s approach to \gls{fr}, with seven stages that capture the most relevant aspects of the frequency control process of a large interconnected power system. It includes communication delays in the signals sent from/to the control center to/from the facilities contracted for regulation, which can significantly impact the \gls{fr} process depending on the magnitude of the delay; for example, for the \gls{ieso}, based on the available data, the average value of the one-way communication delay is about 4 s. However, this is not the case for all the facilities; the medium used to transmit the signals and the physical distance from the sending to the receiving point of the signal proportionally affect the delay. Based on the available data, an extra communication delay $\mathit{CD}$ is included in the aggregated model of \gls{tg} contracted for regulation. The aftereffect of this is an \gls{ace} signal that deviates from zero, and in the worst case scenario, the \gls{fr} signal could worsen the \gls{ace}. For example, it could happen that at time \textit{t}, the system requires a positive regulation action from the \gls{fr} assets; however, due to $\mathit{CD}$, the regulation coming from the facilities could be negative, as a result of calculations based on the state of the system at a time \textit{t} - $\mathit{CD}$; thus, increasing the \gls{ace}. Therefore, delays play an important role in the \gls{fr} control process and thus should be modeled. 

\subsubsection{\gls{ace} Calculation}\label{subsec:IIIA1}
This is the first stage of the frequency control process. The inputs of this block are the actual power in the interconnection $\mathit{NI}_{a}$, and actual frequency $f_{a}$, which are the outputs of the system block; the scheduled power in the interconnection $\mathit{NI}_{s}$ and scheduled frequency $f_{s}$ change over time and are set by the \gls{iso} based on the system's needs; the inadvertent payback $\mathit{IP}$ also changes over time and is set by the \gls{iso}; and the interchange metering error $\mathit{IME}$, which in this case is in manual mode set to -35 MW by the \gls{ieso}. The block has two outputs: the $\mathit{ACE}$ signal and the $\mathit{SR}_{ESS}$ signal, calculated as follows:
\begin{equation}
\resizebox{0.89\hsize}{!}{$
  \mathit{ACE = (NI_{a} - NI_{s}) - 10B(f_{a} - f_{s}) - IME -IP -f(IP)} 
$}\label{eq:ace1}
\end{equation}
\begin{equation}
\resizebox{0.89\hsize}{!}{$
   \mathit{SR_{ESS} = (NI_{a} - NI_{s}) - 10B(f_{a} - f_{s}) - IME -f(IP)}
$}\label{eq:SRegES}
\end{equation}
where the \gls {ba} bias $B$ is equal to $-248.2~\rm{MW/0.1Hz}$ for the \gls{ieso}. 

The $\mathit{f(IP)}$ function in (\ref{eq:ace1}) and (\ref{eq:SRegES}) is introduced to represent the differences between the measured data and the model results, and is associated with the $\mathit{IP}$ signal. This function is modelled here with a \gls{nn}, using three signals as inputs: $\mathit{IP_{t}}$, $\mathit{IP_{t-10}}$, and $\mathit{IP_{t}-IP_{t-10}}$. The \gls{nn} consists of two layers of 48 neurons and 1 neuron, respectively, with a tangent sigmoid as the activation function. The \gls{nn} inputs are normalized between -1 and 1 before entering the training process, and the output is converted back to its real scale. One year data was used to obtain the \gls{nn} model: 80\% for training, 10\% for testing, and 10\% for validation.  
\subsubsection{ACE Filtering} 
\begin{figure}[t!]
    \centering
    \def\svgwidth{0.75\columnwidth}
    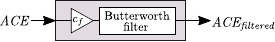
    \caption{Stage II \gls{ace} filtering block.}
    \label{fig:filtered_ACE}
\end{figure}

This corresponds to Stage II in the \gls{fr} process and is presented in Fig.~\ref{fig:filtered_ACE}. The input of this block is the \gls{ace} signal from Stage I, and includes a gain $c_{f}$ and a first order Butterworth filter with a pre-warping frequency $\omega_{0}$. The purpose of this filter is to get rid of fast signal changes, since \glspl{tg} are not able to react to them.
\subsubsection{\gls{agc}} 
\begin{figure}[t!]
    \centering
    \def\svgwidth{0.9\columnwidth}
    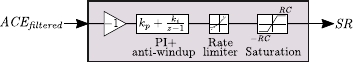
    \caption{Stage III \gls{agc}.}
    \label{fig:AGC}
\end{figure}

It corresponds to Stage III in the \gls{fr} control process, and is depicted on Fig.~\ref{fig:AGC}. The input of this block is the filtered \gls{ace} signal from Stage II, and includes an initial negative gain, because the compensation provided by \gls{fr} should be in the opposite direction to reduce the error. This signal goes into a discrete $\rm PI$ controller with parameters $k_{p}$ and $k_{i}$, and clamping as anti-windup method. The output of the $\rm PI$ block goes through a rate limiter to ensure the generation changes are within limits defined by the \gls{iso} ($\rm{\pm 50~MW/min}$ for the \gls{ieso}), and feeds a saturation block to avoid exceeding the contracted regulation capacity $\mathit{\pm RC}$. The output of this block is the scheduled \gls{fr} signal $\mathit{SR}$, sent to the \glspl{tg} contracted for regulation. 
\subsubsection{Aggregated \gls{tg}} 
\begin{figure}[t!]
    \centering
    \def\svgwidth{0.9\columnwidth}
    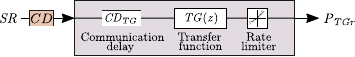
    \caption{Stage IV Aggregated model of \gls{tg} contracted for \gls{fr}.}
    \label{fig:Aggreg_TG}
\end{figure}

It corresponds to Stage IV in the frequency control process, and is illustrated in Fig.~\ref{fig:Aggreg_TG}. Since the real signal available is the aggregated response of all the \glspl{tg} to the $\mathit{SR}$ signal, an aggregated model of these \glspl{tg} is needed; the input of this block is the $\mathit{SR}$ signal from Stage III plus the communication delay $\mathit{CD}$ associated with the \gls{fr} signal, as previously discussed. The delay in the signals in proportional to the physical distance from the control center to the facilities; after analyzing the data available, an extra communication delay $\mathit{CD_{TG}}$ is included in this model (30 s for the \gls{ieso}). The third order transfer function $\mathit{TG(z)}$ represents the action of the \glspl{tg} contracted for regulation, and can be readily estimated from actual measurements. The rate limiter, similar to that in Stage III, ensures the output of this model matches the real data by avoiding unrealistic power changes in the output of the \gls{tg} group. 
\subsubsection{System Model} It represents the primary frequency response of the elements in the system and the power in the tie-lines, and corresponds to Stage VII in the \gls{fr} control process. It has five inputs: the load of the system $\mathit{P_{D}}$; the generation total dispatch $\mathit{P_{GT}}$; and the outputs of the facilities contracted for regulation $\mathit{P_{TGr}}$, $P_{FESS}$, and $P_{BESS}$. This block calculates the actual frequency $\mathit{f_{a}}$ and power in the interconnection $\mathit{NI_{a}}$ at time $t$, as follows:
\begin{multline}
\small
{f_{a}}_{t} = {f_{a}}_{0}+[-\Delta{{P_{D}}_{t}}+\Delta{{P_{GT}}_{t}}+\Delta{{P_{TGr}}_{t}}\\
+\Delta{{P_{FESS}}_{t}}+\Delta{{P_{BESS}}_{t}}]\left[\frac{-1}{B_{EI}}\frac{1}{z-1}-F(z)\right]
\label{eq:freq_a}
\end{multline}
\begin{multline}
\small
{NI_{a}}_{t} = {NI_{a}}_{0}+[-\Delta{{P_{D}}_{t}}+\Delta{{P_{GT}}_{t}}+\Delta{{P_{TGr}}_{t}}\\
+\Delta{{P_{FESS}}_{t}}+\Delta{{P_{BESS}}_{t}}]\left[\frac{{B_{EI}}-{B}}{B_{EI}}\frac{1}{z-1}\right]
\label{eq:NI_a}
\end{multline}
which include the same \gls{ba} bias of (\ref{eq:ace1}) and (\ref{eq:SRegES}), and an \gls{naei} bias $\mathit{B_{EI}}$ calculated from \cite{B_IE_cal}, using available data and a detailed transient stability model. In (\ref{eq:freq_a}), the function $\mathit{F(z)}$ allows obtaining a closer fit between the model results and the measured data.
\subsection{Energy Storage}\label{subsec:IIIB}
\subsubsection{Set-point Calculation} 
\begin{figure}[t!]
 \centering
  \begin{subfigure}[b]{0.45\columnwidth}
    \def\svgwidth{\columnwidth}
    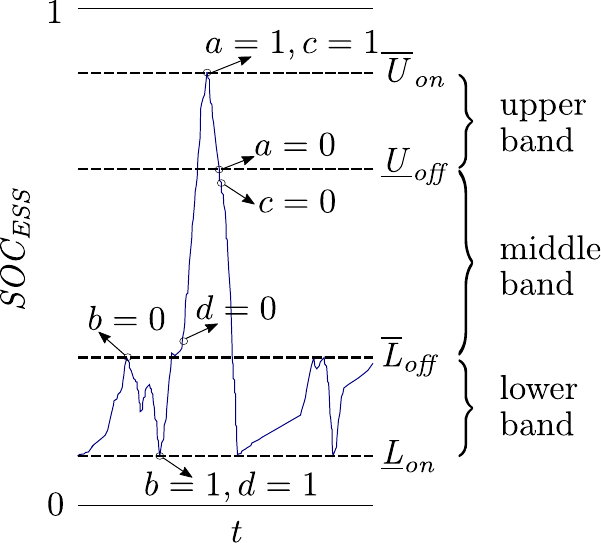
    \caption{}
    \label{fig:ab_index}
  \end{subfigure}
    \begin{subfigure}[b]{0.45\columnwidth}
    \def\svgwidth{\columnwidth}
    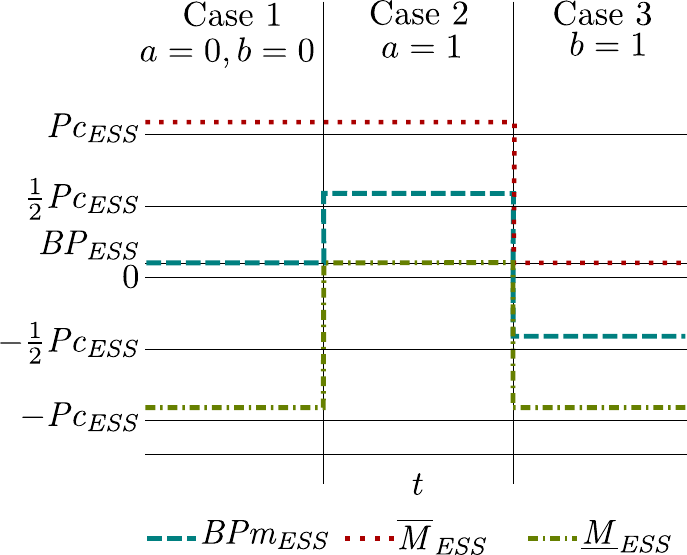
    \caption{}
    \label{fig:min_max}
  \end{subfigure}
  \caption{Relation of (a) parameters $a$, $b$, $c$, and $d$, and (b) $\mathit{\overline{M}_{ESS}}$, $\mathit{\underline{M}_{ESS}}$, and $\mathit{BP_{ESS}}$ to the \gls{SoC} management of the \gls{ess} model.}
  \label{fig:histograms_val2}
 \end{figure}

This is Stage V in the \gls{fr} control process, and includes the calculation of the \gls{sp} signals $\mathit{SP_{FESS}}$ and $\mathit{SP_{BESS}}$ sent from the control center to the \gls{fess} and \gls{bess} facilities, respectively. The calculation of both \gls{sp} signals is as follows, with the inputs changing for each facility:
\begin{equation}
\small
\mathit{SP_{ESS}}=
\resizebox{0.75\hsize}{!}{$
\begin{cases}
\frac{1}{2}(\overline{M}_{ESS}-\underline{M}_{ESS})\frac{min(SR_{ESS},RC)}{RC}+ BPm_{ESS} \\
\quad \quad \quad \forall \ AV_{ESS}=1,  RC\neq 0\ , \ SR_{ESS} \geq=0 \\
\frac{1}{2}(\overline{M}_{ESS}-\underline{M}_{ESS})\frac{max(SR_{ESS},-RC)}{RC} + BPm_{ESS} \\
\quad \quad \quad \forall \ AV_{ESS}=1 ,  RC\neq 0\ , \ SR_{ESS} <0\\
\end{cases}$}
\end{equation}

where $\mathit{AV_{ESS}}$ is the status availability of the facility, $\mathit{\underline{M}_{ESS}}$ and $\mathit{\overline{M}_{ESS}}$ are the minimum and maximum available capacity of the facility, respectively, and $\mathit{BPm_{ESS}}$ is the moving base-point, which is modelled as the fixed base point of the facility $\mathit{BP_{ESS}}$ moving between $\mathit{\underline{M}_{ESS}}$ and $\mathit{\overline{M}_{ESS}}$, as illustrated in Fig.~\ref{fig:min_max}, and containing \gls{SoC} information. Since these signals come from the \gls{ess} facilities, communication delays are considered before they arrive at the control center, as shown in Fig.~\ref{fig:current_model}. In addition, the \gls{fr} capacity limit $\mathit{RC}$, and the $\mathit{SR_{ESS}}$ signal are required in this calculation. The $\mathit{SP_{ESS}}$ signal is in essence a scaled version of the $\mathit{SR_{ESS}}$ signal that takes into account the \gls{SoC} of the \gls{ess} facility reflected through the $\mathit{\overline{M}_{ESS}}$, $\mathit{\underline{M}_{ESS}}$, and $\mathit{BPm_{ESS}}$ signals. 
\begin{figure}[t!]
    \centering
    \def\svgwidth{\columnwidth}
    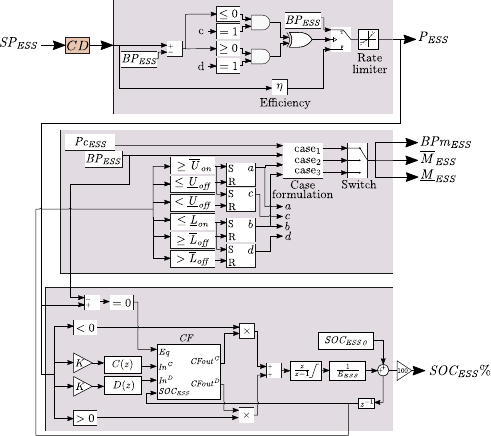
    \caption{Stage VI base \gls{ess} model including \gls{SoC} management.}
    \label{fig:ess_model}
\end{figure}
\subsubsection{\gls{ess} Model} This is Stage VI in the frequency control process and includes the \gls{ess} models of the \gls{bess} and \gls{fess} facilities, as well as their \gls{SoC} management model.  Considering that similar operational data was available for the \gls{bess} and \gls{fess} facilities, similar \gls{ess} models were developed for both facilities, with different parameters and some specific features for each facility.

Fig.~\ref{fig:ess_model} illustrates the base \gls{ess} model, which has two main parts: The first part, which is the output of the \gls{ess} facility in megawatts, depends on the sign of the regulation required from the facility ($\mathit{SP_{ESS}-BP_{ESS}}$), and the \gls{SoC} parameters $c$ and $d$. As shown in Fig.~\ref{fig:ess_model}, the output signal could be equal to $\mathit{BP_{ESS}}$ or to the delayed \gls{sp} signal coming from Stage V, and primarily considers the efficiency of the \gls{ess}. The output signal goes through a ramp rate block with rising slew rate $Rsr$, and falling slew rate $\mathit{Fsr}$. A 100\% charging and discharging efficiency $\eta$ is considered because the actual facility internally compensates the set-point signal taking into account its real efficiency; hence, from the system's perspective, the \gls{ess} acts at full efficiency. 

The second part of the model is the \gls{SoC} management of the facility, and it is divided in two sections. In the first section, $\mathit{BPm_{ESS}}$, $\mathit{\overline{M}_{ESS}}$, and $\mathit{\underline{M}_{ESS}}$, and the four \gls{SoC} parameters $a$, $b$, $c$, and $d$ are calculated. Furthermore, three sections of the \gls{SoC} are considered: a lower band from $\mathit{\underline{L}_{on}}$ to $\mathit{\overline{L}_{off}}$, a middle band from $\mathit{\overline{L}_{off}}$ to $\mathit{\underline{U}_{off}}$, and an upper band from $\mathit{\underline{U}_{off}}$ to $\mathit{\overline{U}_{on}}$. The parameters $a$, $b$, $c$, and $d$ are directly related to these \gls{SoC} bands, as illustrated in Fig. \ref{fig:ab_index} and in the second block of Fig. \ref{fig:ess_model}. If the output of the \gls{ess} causes the \gls{SoC} to reach $\mathit{\underline{L}_{on}}$ ($b=1$) or $\mathit{\overline{U}_{on}}$ ($a=1$), the \gls{ess} starts charging or discharging, respectively, despite the regulation signal. This behaviour continues until the \gls{SoC} reaches $\mathit{\overline{L}_{off}}$ ($b=0$) or $\mathit{\underline{U}_{off}}$ ($a=0$), during the charging/discharging operations, correspondingly. 

After the \gls{SoC} parameters are calculated, and considering the power capacity and base-point of the facility designated for regulation, i.e., $\mathit{Pc_{ESS}}$, and $\mathit{BP_{ESS}}$ signals, respectively, three cases are formulated. Case 1 considers the \gls{SoC} in the middle band of Fig. \ref{fig:ab_index}, where $a=0$ and $b=0$; the second case considers $a=1$, and the third case considers $b=1$. The signals $\mathit{BPm_{ESS}}$, $\mathit{\overline{M}_{{ESS}}}$, and $\mathit{\underline{M}_{{ESS}}}$, illustrated in Fig.~\ref{fig:min_max}, are calculated as follows:
\begin{equation}\label{eq:BPm}
\mathit{BPm_{ESS}} = 
    \begin{cases}
    \mathit{BP_{ESS}} & \forall \quad a=0, b=0\\
    \mathit{BP_{ESS}+{Pc_{ESS}}}  & \forall \quad a=1 \\
    \mathit{BP_{ESS}-{Pc_{ESS}}}  & \forall \quad b=1 \\ 
    \end{cases}    
\end{equation}
\begin{equation}\label{eq:Mx}
\mathit{\overline{M}_{ESS}} =
    \begin{cases}
    \mathit{BP_{ESS}}+{\frac{1}{2}}{Pc_{ESS}} & \forall \quad a=0, b=0\\
    \mathit{BP_{ESS}+{Pc_{ESS}}}   & \forall \quad a=1 \\
    \mathit{BP_{ESS}}  & \forall \quad b=1 \\ 
    \end{cases}       
\end{equation}
\begin{equation}\label{eq:Mn}
\mathit{\underline{M}_{ESS}} =
    \begin{cases}
    \mathit{BP_{ESS}}-{\frac{1}{2}}{Pc_{ESS}}  & \forall \quad a=0, b=0\\
    \mathit{BP_{ESS}}  & \forall \quad a=1 \\
    \mathit{BP_{ESS}-{Pc_{ESS}}}  & \forall \quad b=1 \\ 
    \end{cases}       
\end{equation}

The second section of the \gls{SoC} management model is the \gls{SoC} calculation itself, with the \gls{ess} output power $P_{ESS}$, and the base-point signal $\mathit{BP_{ESS}}$ as inputs, as illustrated in the last block of Fig. \ref{fig:ess_model}. The constant $K$ is the power-to-energy value subject to the sampling-time resolution, which in this case is ${K=-1/3600}$\,~hr/s. Thus, $\mathit{P_{ESS}K}$ is the preliminary charge/discharge energy per sampling-time, which is equivalent to a simplified Coulomb counting \gls{SoC} method~\cite{HANNAN2017834}. Additionally, the model assumes two second order transfer functions to account for the different charging/discharging characteristics $C(z)$ and $D(z)$, respectively, of the \gls{ess} facilities. The outputs of these blocks $\mathit{In^{C}}$ and $\mathit{In^{D}}$, correspondingly, are inputs of the correction factor $\mathit{CF}$ block in  Fig. \ref{fig:ess_model}, which compensates for different charge/discharge energy rates observed in the data provided after $\mathit{\overline{U}_{on}}$ and $\mathit{\underline{L}_{on}}$ are reached, as the charging/discharging slows or speeds up after reaching these limits. The $\mathit{CF}$ block has a charging ($\mathit{CFout^{C}}$) and discharging ($\mathit{CFout^{D}}$) output, which for the \gls{fess} can be defined as follows:
\begin{equation}\label{eq:Correc_factor_ch_FESS}
\resizebox{0.8\hsize}{!}{$
CFout^{C} =
\begin{cases}
In^{C} & \forall \ k_{up}=0,\ k_{dw}=0 \\
In^{C}CF_{eq}  & \forall \ Eq=1,\ (k_{up}=1 \vee k_{dw}=1) \\
In^{C}CF_1^{C}   & \forall \ Eq=0,\ k_{up}=1 \\
In^{C}CF_2^{C}   & \forall \ Eq=0,\ k_{up}=0, \ k_{dw}=1 \\
\end{cases}$}
\end{equation}
\begin{equation}\label{eq:Correc_factor_dch_FESS}
\resizebox{0.7\hsize}{!}{$
CFout^{D} =
\begin{cases}
In^{D} & \forall \ k_{up}=0,\ k_{dw}=0 \\
In^{D}CF_1^{D}   & \forall \ k_{dw}=1 \\
In^{D}CF_2^{D}   & \forall \ k_{up}=1,\ k_{dw}=0 \\
\end{cases}$}
\end{equation}
where $\mathit{CF_{eq}}$, $\mathit{CF_1^{C}}$, $\mathit{CF}_2^{C}$ are estimated parameters that multiply $\mathit{In^{C}}$, according to the conditions shown in (\ref{eq:Correc_factor_ch_FESS}). The input $\mathit{Eq}$ takes the value of $1$ when the $P_{FESS}$ is equal to $\mathit{BPm_{FESS}}$. Furthermore, $\mathit{CF_1^{D}}$ and $\mathit{CF}_2^{D}$ are estimated parameters, which multiply $\mathit{In^{D}}$, according to the conditions in (\ref{eq:Correc_factor_dch_FESS}). The variables $k_{up}$ and $k_{dw}$ are the outputs of set-reset flip-flops defined as follows:
\begin{equation}\label{eq:kup}
{k_{up}}_{t} = {S_{up}}_{t}+{k_{up}}_{t-1}({R_{up}}_{t}\times -1)
\end{equation}
\begin{equation}\label{eq:kdw}
{k_{dw}}_t = {S_{dw}}_t+{k_{dw}}_{t-1}({R_{dw}}_{t}\times -1)
\end{equation}
where
\begin{equation}\label{eq:Sup}
{S_{up}}_t = 1 \ \forall \ SOC_{ESS}\geq ks_{up}  
\end{equation}
\begin{equation}\label{eq:Rup}
{R_{up}}_t = 1 \ \forall \ SOC_{ESS}\leq kr_{up}  
\end{equation}
\begin{equation}\label{eq:Sdw}
{S_{dw}}_t = 1 \ \forall \ SOC_{ESS}\leq ks_{dw} 
\end{equation}
\begin{equation}\label{eq:Rdw}
{R_{dw}}_t = 1 \ \forall \ SOC_{ESS}\geq kr_{dw}  
\end{equation}
and $ks_{up}$ and $kr_{up}$ are estimated parameters that represent the values of the \gls{SoC} that cause the set ${S_{up}}_{t}$ and reset ${R_{up}}_{t}$ signals of the flip-flop ${k_{up}}_{t}$ to become $1$. Likewise, $ks_{dw}$ and $kr_{dw}$ are estimated parameters associated with the \gls{SoC} values that activates the set (${S_{dw}}_{t}$) and reset (${R_{dw}}_{t}$) signals of the flip-flop  ${k_{dw}}_{t}$, correspondingly.

For the case of \gls{bess}, the $\mathit{CF}$ block can be defined as follows:
\begin{equation}\label{eq:Correc_factor_ch_BESS}
\resizebox{0.8\hsize}{!}{$
CFout^{C} =
\begin{cases}
In^{C} & \forall \ k_{up}=0,\ k_{dw}=0 \\
In^{C}\left[\frac{1}{1+SOC_{BESS}}\right]CF_1^{C}   & \forall \ k_{up}=1 \\
In^{C}\left[1+SOC_{BESS}\right]CF^{C2}   & \forall \ k_{up}=0,\ k_{dw}=1 \\
\end{cases}$}
\end{equation}
\begin{equation}\label{eq:Correc_factor_dch_BESS}
\resizebox{0.8\hsize}{!}{$
CFout^{D} =
\begin{cases}
In^{D} & \forall \ k_{up}=0,\ k_{dw}=0 \\
In^{D}\left[1+SOC_{BESS}\right]CF^{D1}   & \forall \ k_{dw}=1 \\
In^{D}\left[\frac{1}{1+SOC_{BESS}}\right]CF^{D2}   & \forall \ k_{up}=1,\ k_{dw}=0 \\
\end{cases}$}
\end{equation}
where $\mathit{CF_1^{C}}$, $\mathit{CF}_2^{C}$, and $\mathit{CF_1^{D}}$, $\mathit{CF}_2^{D}$ are estimated parameters that multiply a function of the \gls{SoC}, and the inputs $\mathit{In^{C}}$ and $\mathit{In^{D}}$, respectively. The variables $k_{up}$ and $k_{dw}$ in (\ref{eq:Correc_factor_ch_BESS}) and (\ref{eq:Correc_factor_dch_BESS}) are the same as in (\ref{eq:kup}) and (\ref{eq:kdw}). 

The transfer functions $C(z)$ and $D(z)$ both have a parallel comparison block (binary variables) to operate in either mode, which are related to the value of $\mathit{P_{ESS}}$; these binary variables are multiplied by $\mathit{CFout^{C}}$ and $\mathit{CFout^{D}}$, as shown in Fig. \ref{fig:ess_model}. Finally, the estimated corrected energy for the sampling interval is integrated and divided by the \gls{ess} energy capacity $\mathit{E_{ESS}}$ and added to the initial \gls{SoC} value to obtain the estimated \gls{SoC} output at time $t$. This value is later multiplied by 100\% to obtain $\mathit{SOC_{ESS}\%}$ at time $t$. The \gls{SoC} empirical model proposed in this paper is derived by analyzing operational data for an actual \gls{fess} and a \gls{bess} used for \gls{fr} by the \gls{ieso}. Note that the \gls{SoC} model does not consider degradation or cell failure which will impact the output; further data would be required to model such effects. 

Currently, the base-point signal $\mathit{BP_{ESS}}$ for the \gls{ieso} is zero or a value close to zero, set in each \gls{ess} facility. However, the control center may replace the $\mathit{BP_{ESS}}$ with the dispatch from the energy market for facilities that are able to simultaneously participate in both markets. In such a situation, the regulation required from the \gls{ess} facilities would be in addition to the base-point signal ($\mathit{SP_{ESS}-BP_{ESS}}$), as currently is the case for \glspl{tg}. 
\setlength{\tabcolsep}{5pt}
\begin{table}[t!]
\centering
\caption{Model Parameters.}
\resizebox{1\columnwidth}{!}{%
	\begin{tabular}{lclc}
				\hhline{====}
		\multicolumn{4}{c}{{\rule{0pt}{1ex} \textbf{Stage II: ACE filtering}}}\\
		\hhline{====}
		\multicolumn{1}{c}{Parameter} & \multicolumn{1}{c}{Value} & \multicolumn{1}{c}{Parameter} & \multicolumn{1}{c}{Value}\\
		\hhline{----}
		\multicolumn{1}{l}{\rule{0pt}{1ex}$c_{f}$}	&  \multicolumn{1}{c}{$0.974$} 	&  \multicolumn{1}{l}{\rule{0pt}{1ex}$\omega_{0}$~{\scriptsize [rad/s]}} &	\multicolumn{1}{c}{$0.097$}\\
		\hhline{====}
		\multicolumn{4}{c}{{\rule{0pt}{1ex} \textbf{Stage III: \gls{agc}}}}\\
		\hhline{====}
		\multicolumn{1}{c}{Parameter} & \multicolumn{1}{c}{Value} & \multicolumn{1}{c}{Parameter} & \multicolumn{1}{c}{Value}\\
		\hhline{----}
		\multicolumn{1}{l}{\rule{0pt}{1ex}$k_{i}$}	&  \multicolumn{1}{c}{$0.022$} 	&  \multicolumn{1}{l}{\rule{0pt}{1ex}$k_{p}$} &	\multicolumn{1}{c}{$0.42$}\\
		\hhline{====}
		\multicolumn{4}{c}{{\rule{0pt}{1ex} \textbf{Stage IV: Aggregated model of \gls{tg} contracted for \gls{fr}}}}\\
		\hhline{====}
		\multicolumn{1}{c}{Parameter} & \multicolumn{1}{c}{Value} & &\\
		\hhline{----}
		\multicolumn{1}{l}{\rule{0pt}{1ex}$\mathit{TG(z)}$}	&  \multicolumn{1}{c}{$\frac{3.45z^{2}+1.58}{3.78z^{3}+1.47z^{2}}$} \\
		\hhline{====}
		\multicolumn{4}{c}{{\rule{0pt}{1ex} \textbf{Stage VI: \gls{ess} models}}}\\
		\hhline{====}
		\multicolumn{2}{c|}{{\rule{0pt}{1ex} \gls{fess} Model}} & \multicolumn{2}{c}{{ \gls{bess} Model}} \\
		\hhline{----}
		\multicolumn{1}{c}{Parameter} & \multicolumn{1}{c|}{Value} & \multicolumn{1}{c}{Parameter} & \multicolumn{1}{c}{Value}\\
		\hhline{----}
		\multicolumn{1}{l}{\rule{0pt}{0ex} $\mathit{Rsr}${\scriptsize [MW]}}         & \multicolumn{1}{c|}{$0.6$}       &  \multicolumn{1}{l}{\rule{0pt}{1ex} $\mathit{Rsr}${\scriptsize [MW]}}     & \multicolumn{1}{c}{$1.28$}\\
		\multicolumn{1}{l}{\rule{0pt}{0ex} $\mathit{Fsr}${\scriptsize [MW]}}         & \multicolumn{1}{c|}{$-0.6$}      &  \multicolumn{1}{l}{\rule{0pt}{1ex} $\mathit{Fsr}${\scriptsize [MW]}}     & \multicolumn{1}{c}{$-1.28$}\\
		\multicolumn{1}{l}{\rule{0pt}{0ex} $\mathit{\overline{U}_{on}}$}                           & \multicolumn{1}{c|}{$1$}    &  \multicolumn{1}{l}{\rule{0pt}{1ex} $\mathit{\overline{U}_{on}}$}                       & \multicolumn{1}{c}{$0.885$}\\
		\multicolumn{1}{l}{\rule{0pt}{0ex} $\mathit{\underline{U}_{off}}$}                          & \multicolumn{1}{c|}{$0.75$}    &  \multicolumn{1}{l}{\rule{0pt}{1ex} $\mathit{\underline{U}_{off}}$}                      & \multicolumn{1}{c}{$0.885$}\\
		\multicolumn{1}{l}{\rule{0pt}{0ex} $\mathit{\overline{L}_{off}}$}                          & \multicolumn{1}{c|}{$0.25$}    &  \multicolumn{1}{l}{\rule{0pt}{1ex} $\mathit{\overline{L}_{off}}$}                      & \multicolumn{1}{c}{$0.125$}\\
		\multicolumn{1}{l}{\rule{0pt}{0ex} $\mathit{\underline{L}_{on}}$}                           & \multicolumn{1}{c|}{$0$}       &  \multicolumn{1}{l}{\rule{0pt}{1ex} $\mathit{\underline{L}_{on}}$}                       & \multicolumn{1}{c}{$0.125$} \\
		\multicolumn{1}{l}{\rule{0pt}{0ex} $K$}                          & \multicolumn{1}{c|}{$-1/3600$}    &  \multicolumn{1}{l}{\rule{0pt}{1ex} $K$}                      & \multicolumn{1}{c}{$-1/3600$}\\
		\multicolumn{1}{l}{\rule{0pt}{1ex} $C(z)$}                             & \multicolumn{1}{c|}{$\frac{4.23z^{2}+1.06z+2.8}{z^{2}+0.57z+0.42}$} &  \multicolumn{1}{l}{\rule{0pt}{1ex} $C(z)$} & \multicolumn{1}{c}{$\frac{0.16z^{2}+4.11z+7.52}{z^{2}+0.93z+0.54}$}\\
		\multicolumn{1}{l}{\rule{0pt}{1ex} $D(z)$} & \multicolumn{1}{c|}{$\frac{5.57z^{2}+6.72z+3.02}{z^{2}+0.72z+0.81}$} &  \multicolumn{1}{l}{\rule{0pt}{1ex} $D(z)$} & \multicolumn{1}{c}{$\frac{2.88z^{2}+3.78z+4.57}{z^{2}+0.48z+0.55}$}\\
		\multicolumn{1}{l}{\rule{0pt}{0ex} $\mathit{CF_1^{C}}$}                           & \multicolumn{1}{c|}{$1.22$}    &  \multicolumn{1}{l}{\rule{0pt}{1ex} $\mathit{CF_1^{C}}$}                           & \multicolumn{1}{c}{$0.014$}\\
		\multicolumn{1}{l}{\rule{0pt}{0ex} $\mathit{CF}_2^{C}$}                           & \multicolumn{1}{c|}{$0.99$}    &  \multicolumn{1}{l}{\rule{0pt}{1ex} $\mathit{CF}_2^{C}$}                           & \multicolumn{1}{c}{$0.50$}\\
		\multicolumn{1}{l}{\rule{0pt}{0ex} $\mathit{CF_1^{D}}$}                           & \multicolumn{1}{c|}{$1.19$}    &  \multicolumn{1}{l}{\rule{0pt}{1ex} $\mathit{CF_1^{D}}$}                          & \multicolumn{1}{c}{$0.51$}\\
		\multicolumn{1}{l}{\rule{0pt}{0ex} $\mathit{CF}_2^{D}$}                           & \multicolumn{1}{c|}{$1.10$}    &  \multicolumn{1}{l}{\rule{0pt}{1ex} $\mathit{CF}_2^{D}$}                           & \multicolumn{1}{c}{$0.64$}\\
		\multicolumn{1}{l}{\rule{0pt}{0ex} $\mathit{CF_{eq}}$}                          & \multicolumn{1}{c|}{$0.35$}    &  \multicolumn{1}{l}{\rule{0pt}{1ex} $\mathit{CF_{eq}}$}                          & \multicolumn{1}{c}{$-$}\\
		\multicolumn{1}{l}{\rule{0pt}{0ex} $ks_{up}$}                      & \multicolumn{1}{c|}{$1$}    &  \multicolumn{1}{l}{\rule{0pt}{1ex} $ks_{up}$}                     & \multicolumn{1}{c}{$0.885$}\\
		\multicolumn{1}{l}{\rule{0pt}{0ex} $kr_{up}$}                      & \multicolumn{1}{c|}{$0.55$}    &  \multicolumn{1}{l}{\rule{0pt}{1ex} $kr_{up}$}                      & \multicolumn{1}{c}{$0.884$}\\
		\multicolumn{1}{l}{\rule{0pt}{0ex} $ks_{dw}$}                      & \multicolumn{1}{c|}{$0$}    &  \multicolumn{1}{l}{\rule{0pt}{1ex} $ks_{dw}$}                      & \multicolumn{1}{c}{$0.125$}\\
		\multicolumn{1}{l}{\rule{0pt}{0ex} $kr_{dw}$}                      & \multicolumn{1}{c|}{$0.60$}    &  \multicolumn{1}{l}{\rule{0pt}{1ex} $kr_{dw}$}                    & \multicolumn{1}{c}{$0.144$}\\
		\hhline{----}
		\multicolumn{4}{c}{{\rule{0pt}{1ex} \textbf{Stage VII: System model}}}\\
		\hhline{----}
		\multicolumn{1}{c}{Parameter} & \multicolumn{3}{c}{Value} \\
		\hhline{----}
		\multicolumn{1}{l}{\rule{0pt}{1ex}$\mathit{F(z)}$}	&  \multicolumn{3}{l}{$\frac{0.51z^{5}-1.97z^{4}+2.31z^{3}+2.93z^{2}-1.18z-2.73}{z^{5}-0.45z^{4}-0.77z^{3}-0.23z^{2}+0.6z-0.12}e^{-5}$} \\
		\hhline{====}
	\end{tabular}}
	\label{table:parvalues}
\end{table}
\section{Validation of Proposed \gls{fr} model on \gls{ops}}\label{sec:IV}
All the stages in the proposed \gls{fr} model shown in Fig. \ref{fig:current_model} were validated using information provided by the \gls{ieso}, which included a \gls{dsatools} model of the \gls{naei}, \gls{ops} data, and data from a 2 MW/0.5 MWh \gls{fess}, and a 4 MW/2.76 MWh \gls{bess} used for \gls{fr} by the \gls{ieso}. All the parameter values of the proposed model determined for the \gls{ops} are presented in Table \ref{table:parvalues}. 
\subsection{Test System Validation}\label{subsec:IVA}
The first step in the validation process of the proposed \gls{fr} model is the validation of frequency response of the \gls{dsatools} model, which is referred here as the test system, against real data. This test system is a reduced representation of the \gls{naei}, with a detailed representation of the \gls{ops} and a combination of detailed and equivalent aggregated models, depending on their impact and electrical distances, of the external area.

For validation purposes, seven scenarios capturing the frequency response of the \gls{ops} were selected based on regulation signal changes. The load changes for the \gls{naei} were determined based on the measured frequency profile and load changes in the \gls{ops}; a frequency response value of -2760 MW/0.1 Hz for the \gls{naei}; and the initial powers of the load, generation, and the interconnection for each scenario. The difference between the total expected load change for the \gls{naei} and the total change in the loads in the \gls{ops} was proportionally distributed to all the loads in the rest of the interconnection. Furthermore, three hydroelectric generators in the \gls{ops}, with enough capacity to follow the regulation signal, were selected for the provision of \gls{fr}; the regulation signal was equally divided and sent to these generators. Because of the large number of changes per second, the maximum possible simulation time using the \gls{dsatools} software is 102 seconds. Since the simulation period is less than two minutes and the dispatch changes every 5 minutes, it was assumed that the generator active powers were fixed for each scenario. 

The \gls{rmse} and \gls{mae} of \gls{ace} for the seven cases on March 20, 2018 are presented in Table \ref{table:FR_val}, which can be considered acceptable. Thus, this validated \gls{ops} test system was used next to validate the system block of the proposed \gls{fr} model. The difference between the measured data and test system results could be due to the following reasons:
\begin{itemize}
    \item The value of frequency response used to determine the power changes in the interconnection was assumed to be fixed for all the scenarios. However, with changes in loads, dispatched generators, and connected renewable generators, this value may change depending on the scenario. 
    \item Since the load models are voltage dependent, and their parameters remain fixed during the simulation, while in the actual system these vary, the modelled load powers may not be the same as in the actual system. 
    \item The dynamic file was modified to obtain a more realistic response from the system. However, these modifications may not be an exact representation of the day selected for simulations.
\end{itemize}
\setlength{\tabcolsep}{4pt}
\begin{table}[t!]
\small
\centering
\caption{Validation of the frequency response of the \gls{dsatools} model.}
\resizebox{\columnwidth}{!}{%
  \begin{tabular}{cccccc}
		\hhline{======}
		\multicolumn{1}{c}{} & \multicolumn{1}{c}{\gls{rmse} [Hz]} & \multicolumn{1}{c}{\gls{mae} [Hz]} & \multicolumn{1}{c}{} &  \multicolumn{1}{c}{\gls{rmse} [Hz]} & \multicolumn{1}{c}{\gls{mae} [Hz]}\\
		\hhline{======}
		\multicolumn{1}{l}{\rule{0pt}{1ex} \textbf{Case 1}}	&  \multicolumn{1}{c}{$0.00101$} &	\multicolumn{1}{c}{$0.00082$}	& \multicolumn{1}{l}{\rule{0pt}{1ex} \textbf{Case 5}}	 &  \multicolumn{1}{c}{$0.00279$} 	&	\multicolumn{1}{c}{$0.00225$}\\
		\hhline{------}
		\multicolumn{1}{l}{\rule{0pt}{1ex} \textbf{Case 2}}	&  \multicolumn{1}{c}{$0.00103$} &	\multicolumn{1}{c}{$0.00084$}	& \multicolumn{1}{l}{\rule{0pt}{1ex} \textbf{Case 6}}	&	\multicolumn{1}{c}{$0.00183$} &  \multicolumn{1}{c}{$0.00139$} 	\\
		\hhline{------}
		\multicolumn{1}{l}{\rule{0pt}{1ex} \textbf{Case 3}}	&  \multicolumn{1}{c}{$0.00186$} &	\multicolumn{1}{c}{$0.00129$}	& \multicolumn{1}{l}{\rule{0pt}{1ex} \textbf{Case 7}}	&	\multicolumn{1}{c}{$0.00470$} &  \multicolumn{1}{c}{$0.00359$} 	\\
		\hhline{------}
		\multicolumn{1}{l}{\rule{0pt}{1ex} \textbf{Case 4}}	&  \multicolumn{1}{c}{$0.00592$}  &	\multicolumn{1}{c}{$0.00512$}	&	 &   	&	\\
		\hhline{======}
	\end{tabular}}
	\label{table:FR_val}
\end{table}
\subsection{\gls{fr} Model Validation}\label{subsec:IVB}
 \begin{figure}[t!]
 \centering
  \begin{subfigure}[b]{0.45\columnwidth}
    \def\svgwidth{\columnwidth}
    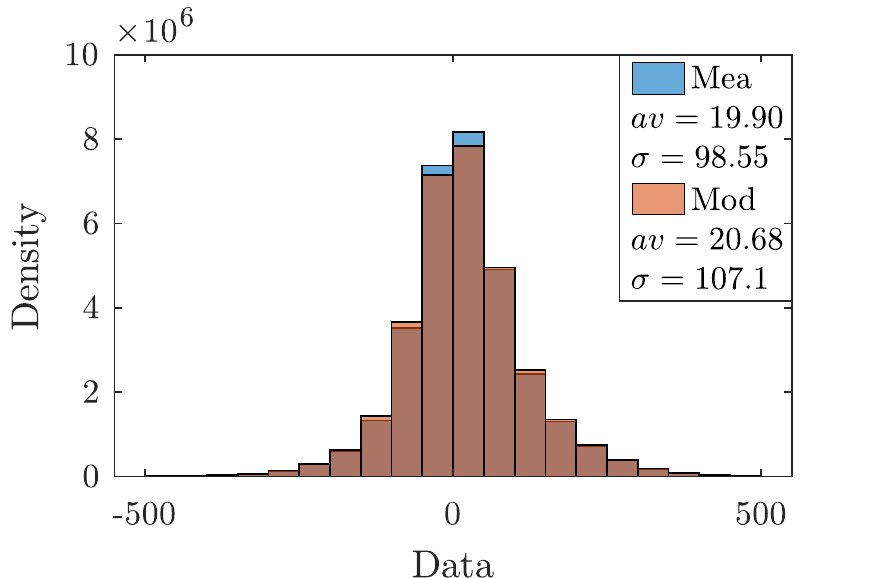
    \caption{$\mathit{ACE}$}
    \label{fig:ACE_his}
  \end{subfigure}
  \begin{subfigure}[b]{0.45\columnwidth}
    \def\svgwidth{\columnwidth}
    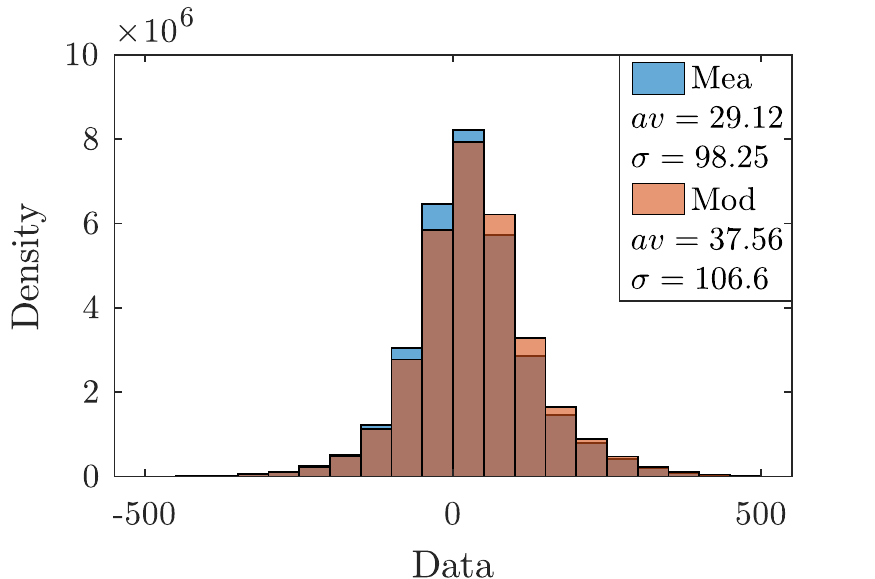
    \caption{$\mathit{SR_{ESS}}$}
    \label{fig:Sreg_ES_his}
  \end{subfigure}
  \begin{subfigure}[b]{0.45\columnwidth}
    \def\svgwidth{\columnwidth}
    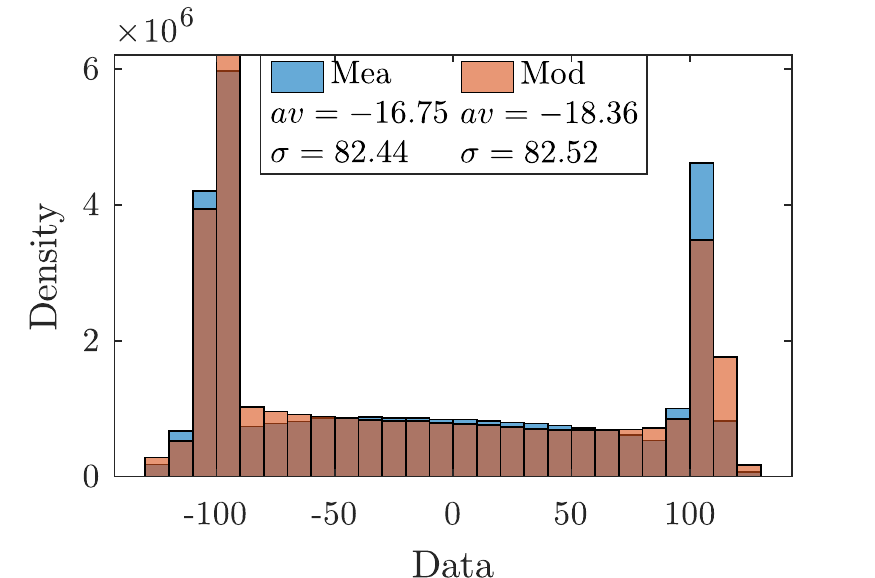
    \caption{$\mathit{SR}$}
    \label{fig:filter_AGC_his}
  \end{subfigure}
    \begin{subfigure}[b]{0.45\columnwidth}
    \def\svgwidth{\columnwidth}
    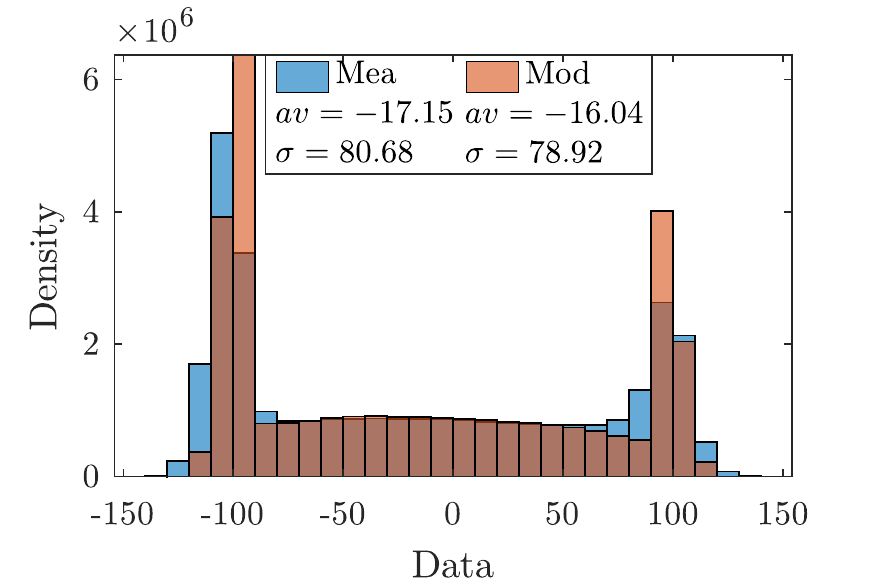
    \caption{$\mathit{P_{TGr}}$}
    \label{fig:filter_TGreg_his}
  \end{subfigure}
  \caption{Histograms of the measured data and model results for Stages I, III and IV.}
  \label{fig:histograms_val1}
 \end{figure}
All the stages in the proposed \gls{fr} model are validated in this subsection. The data made available by the \gls{ieso} corresponds to all the signals associated with the proposed \gls{fr} model (Fig. \ref{fig:current_model}), which has two sections: the bulk system, which includes Stages I to IV, and Stage VII, and the \gls{ess} section, which includes Stage V and VI. For the validation of Stage I to Stage IV, one year of data (April-2018 to March-2019) was used. Fig. \ref{fig:histograms_val1} shows histograms comparing the measured data and the model results for the $\mathit{ACE}$, $\mathit{SR_{ESS}}$, $\mathit{SR}$, and $\mathit{P_{TGr}}$ signals in MW.
\setlength{\tabcolsep}{4pt}
\begin{table}[t!]
\small
\centering
\caption{Estimation errors for $f_{a}$ and $NI_{a}$.}
\resizebox{\columnwidth}{!}{%
  \begin{tabular}{ccccc}
		\hhline{=====}
		\multicolumn{1}{c}{} & \multicolumn{2}{c}{Frequency} & \multicolumn{2}{c}{$\mathit{NI}_{a} \  (523-731  \ \rm{MW})$}\\
		\hhline{=====}
		\multicolumn{1}{c}{} & \multicolumn{1}{c}{\gls{rmse} [Hz]} & \multicolumn{1}{c}{\gls{mae} [Hz]} & \multicolumn{1}{c}{\gls{rmse} [MW]} & \multicolumn{1}{c}{\gls{mae} [MW]}\\
		\hhline{=====}
		\multicolumn{1}{l}{\rule{0pt}{1ex} \textbf{Case 1}}	&  \multicolumn{1}{c}{$2.7329e^{-04}$} 	&	\multicolumn{1}{c}{$2.4423e^{-04}$} &  \multicolumn{1}{c}{$4.4963$} 	&	\multicolumn{1}{c}{$4.0506$}\\
		\hhline{-----}
		\multicolumn{1}{l}{\rule{0pt}{1ex} \textbf{Case 2}}	&  \multicolumn{1}{c}{$0.9063e^{-04}$} 	&	\multicolumn{1}{c}{$0.8111e^{-04}$} &  \multicolumn{1}{c}{$5.1326$} 	&	\multicolumn{1}{c}{$4.1198$}\\
		\hhline{-----}
		\multicolumn{1}{l}{\rule{0pt}{1ex} \textbf{Case 3}}	&  \multicolumn{1}{c}{$1.8753e^{-04}$} 	&	\multicolumn{1}{c}{$1.5642e^{-04}$} &  \multicolumn{1}{c}{$3.9204$} 	&	\multicolumn{1}{c}{$3.2470$}\\
		\hhline{-----}
		\multicolumn{1}{l}{\rule{0pt}{1ex} \textbf{Case 4}}	&  \multicolumn{1}{c}{$3.2162e^{-04}$} 	&	\multicolumn{1}{c}{$2.5331e^{-04}$} &  \multicolumn{1}{c}{$15.953$} 	&	\multicolumn{1}{c}{$14.161$}\\
		\hhline{-----}
		\multicolumn{1}{l}{\rule{0pt}{1ex} \textbf{Case 5}}	&  \multicolumn{1}{c}{$1.9111e^{-04}$} 	&	\multicolumn{1}{c}{$1.4437e^{-04}$} &  \multicolumn{1}{c}{$5.1016$} 	&	\multicolumn{1}{c}{$3.8430$}\\
		\hhline{-----}
		\multicolumn{1}{l}{\rule{0pt}{1ex} \textbf{Case 6}}	&  \multicolumn{1}{c}{$1.0038e^{-04}$} 	&	\multicolumn{1}{c}{$0.7776e^{-04}$} &  \multicolumn{1}{c}{$4.0183$} 	&	\multicolumn{1}{c}{$3.3039$}\\
		\hhline{-----}
		\multicolumn{1}{l}{\rule{0pt}{1ex} \textbf{Case 7}}	&  \multicolumn{1}{c}{$0.3653e^{-04}$} 	&	\multicolumn{1}{c}{$3.0986e^{-04}$} &  \multicolumn{1}{c}{$5.3813$} 	&	\multicolumn{1}{c}{$4.1019$}\\
		\hhline{=====}
	\end{tabular}}
	\label{table:fa_error}
\end{table}
\begin{figure}[t!]
    \centering
    \fontsize{5pt}{10pt}\selectfont%
    \def\svgwidth{0.92\columnwidth}
    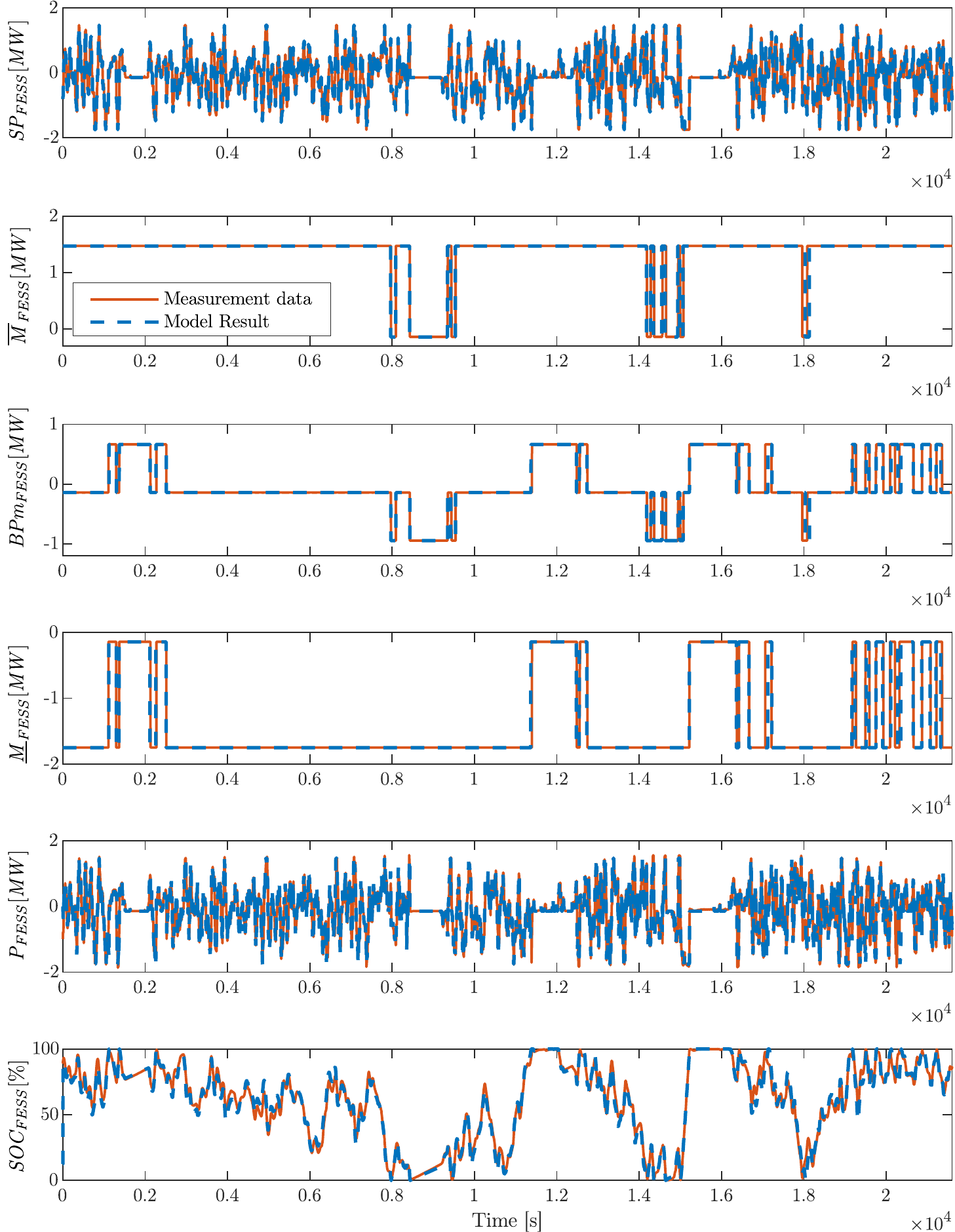
    \caption{\gls{fess} validation results for February 26, 2020.}
    \label{fig:FESS_val}
\end{figure}
\begin{figure}[t!]
    \centering
    \fontsize{5pt}{10pt}\selectfont%
    \def\svgwidth{0.92\columnwidth}
    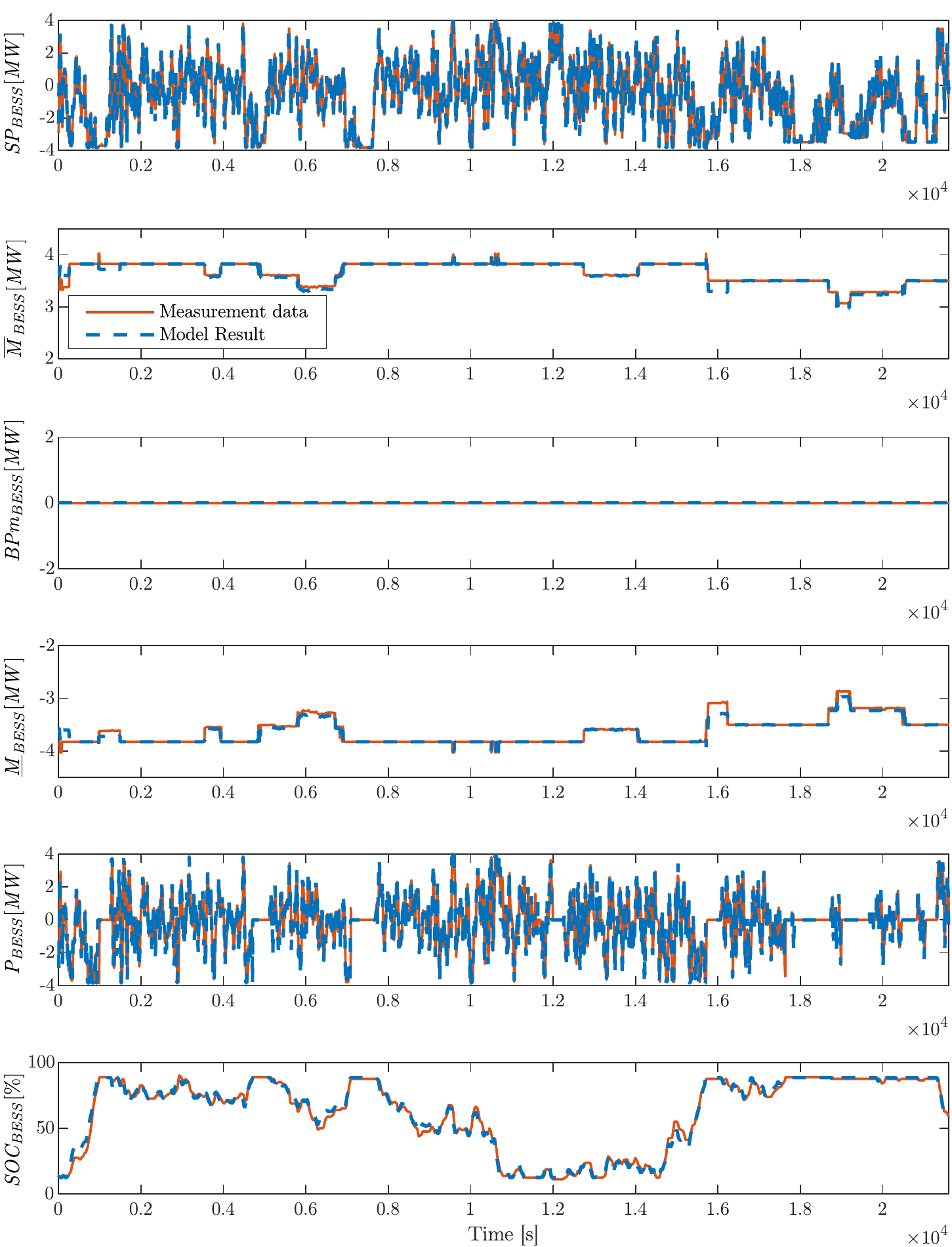
    \caption{\gls{bess} validation results for January 7, 2020.}
    \label{fig:BESS_val}
\end{figure}

After the validation of the frequency response of the test system in the previous subsection, the Stage VII Simulink model is validated against the test system. Since the measured data of load changes in the neighboring interconnected areas is not available with the same resolution, and considering each \gls{ba} is mainly responsible for compensating the load changes within its own area, load variations occurring only in the \gls{ba} of interest are considered here. The same cases used for the validation of the test system were used here, with the difference of no load changes considered in the other \glspl{ba} within the \gls{naei}. The \gls{rmse} and \gls{mae} of \gls{ace} for the seven cases are presented in Table \ref{table:fa_error}. The minimum and maximum $NI_{s}$ values among the seven cases are 523 and 731 MW, respectively, presented for comparison purposes. 

The \gls{ess} section of the proposed \gls{fr} model is validated next. Thus, both Stage V and Stage VI of the \gls{fr} control process are validated for the \gls{fess} and \gls{bess} using one day of data for each facility. The target signals are $\mathit{SP_{ESS}}$ from Stage V, and $\mathit{BPm_{ESS}}$, $\overline{M}_{ESS}$, $\underline{M}_{ESS}$, $P_{ESS}$ and $\mathit{SOC_{ESS}}$ from Stage VI. Simulation results comparing the measured data and the model results for the signals in these two stages are presented in Fig. \ref{fig:FESS_val} and Fig. \ref{fig:BESS_val} for \gls{fess} and \gls{bess}, respectively. In these figures, it can be observed that the signals from the proposed \gls{fr} model closely follow the measurement data from the \gls{ess} facilities. For visualization purposes, only a time span of six hours is presented in these figures, while Table \ref{table:ES_val3} presents the \gls{mae} and \gls{rmse} of the \gls{fess} and \gls{bess} models for a period of one day, for all the signals in the \gls{sp} calculation and \gls{ess} model blocks. 
\setlength{\tabcolsep}{2pt}
\begin{table}[t!]
\small
\centering
\caption{Estimation errors for \gls{fess} and \gls{bess} models for a day.}
\resizebox{\columnwidth}{!}{%
    \begin{tabular}{cccccccc}
    \hhline{~~======}
     \multicolumn{2}{c}{} & \multicolumn{1}{c}{\rule{0pt}{1ex} $\mathit{SP_{ESS}}$}	& \multicolumn{1}{c}{\rule{0pt}{1ex} $\overline{M}_{ESS}$} & \multicolumn{1}{c}{\rule{0pt}{1ex} $\mathit{BPm_{ESS}}$} & \multicolumn{1}{c}{\rule{0pt}{1ex}                                                                $\underline{M}_{ESS}$}	& \multicolumn{1}{c}{\rule{0pt}{1ex} $P_{ESS}$}	  & \multicolumn{1}{c}{\rule{0pt}{1ex} $\mathit{SOC_{ESS}}$} \\
    \multicolumn{2}{c}{} & \multicolumn{1}{c}{\rule{0pt}{1ex} [MW]}	& \multicolumn{1}{c}{\rule{0pt}{1ex} [MW]} & \multicolumn{1}{c}{\rule{0pt}{1ex} [MW]} & \multicolumn{1}{c}{\rule{0pt}{1ex}                                                                [MW]}	& \multicolumn{1}{c}{\rule{0pt}{1ex} [MW]}	  & \multicolumn{1}{c}{\rule{0pt}{1ex} $[\%]$} \\
     \hhline{========}
      \multirow{1}{*}{\gls{fess} } & \multirow{1}{*}{MAE} & \multicolumn{1}{|c}{$0.252$} & \multicolumn{1}{c}{$0.041$} & \multicolumn{1}{c}{$0.064$} & \multicolumn{1}{c}{$0.098$} & \multicolumn{1}{c}{$0.204$} & \multicolumn{1}{c}{$5.072$}\\ 
     \multirow{1}{*}{ ($\pm$ 2MW)} & \multirow{1}{*}{RMSE} & \multicolumn{1}{|c}{$0.425$} & \multicolumn{1}{c}{$0.248$} & \multicolumn{1}{c}{$0.226$} & \multicolumn{1}{c}{$0.390$} & \multicolumn{1}{c}{$0.372$} & \multicolumn{1}{c}{$7.450$} \\ 
     \hhline{--------}
      \multirow{1}{*}{\gls{bess}} & \multirow{1}{*}{MAE} & \multicolumn{1}{|c}{$0.479$} & \multicolumn{1}{c}{$0.045$} & \multicolumn{1}{c}{$0.000$} & \multicolumn{1}{c}{$0.045$} & \multicolumn{1}{c}{$0.330$} & \multicolumn{1}{c}{$1.721$}\\ 
      \multirow{1}{*}{($\pm$4MW)} & \multirow{1}{*}{RMSE} & \multicolumn{1}{|c}{$0.711$} & \multicolumn{1}{c}{$0.102$} & \multicolumn{1}{c}{$0.000$} & \multicolumn{1}{c}{$0.102$} & \multicolumn{1}{c}{$0.609$} & \multicolumn{1}{c}{$2.523$} \\ 
    \hhline{========}
    \end{tabular}}
    \label{table:ES_val3}
\end{table}
\subsection{Communication Delays}\label{subsec:IVC}
\setlength{\tabcolsep}{3pt}
\begin{table}[t!]
\small
\centering
\caption{Impact of communication delays for 100-day period.}
  \begin{tabular}{lcc}
		\hhline{===}
		\multirow{1}{*}{\rule{0pt}{1ex} \textbf{Cases}} & \multicolumn{1}{c}{\gls{rmse} [MW]} & \multicolumn{1}{c}{\gls{mae} [MW]} \\
	    \hhline{===}
		\multirow{1}{*}{\rule{0pt}{1ex} Current delay}	&  \multirow{1}{*}{$87.55$}	&	\multirow{1}{*}{$56.91$}  \\
		\hhline{---}
		\multirow{1}{*}{\rule{0pt}{1ex} Half of current delay}	&  \multirow{1}{*}{$84.14 $}	&	\multirow{1}{*}{$52.50 $}\\
		\hhline{---}
		\multirow{1}{*}{\rule{0pt}{1ex} No delay}	&  \multirow{1}{*}{$80.64 $}	&	\multirow{1}{*}{$47.12 $}	 \\
	    \hhline{===}
	\end{tabular}
	\label{table:impact_delay}
\end{table}
\setlength{\tabcolsep}{2pt}
\begin{table}[t!]
\small
\centering
\caption{Impact of \gls{SoC} model for 100-day period.}
  \begin{tabular}{lcc}
		\hhline{===}
		\multirow{1}{*}{\rule{0pt}{1ex} \textbf{Cases}} & \multicolumn{1}{c}{\gls{rmse} [MW]}  & \multicolumn{1}{c}{\gls{mae} [MW]} \\
	    \hhline{===}
	    \multirow{1}{*}{\rule{0pt}{1ex} Current \gls{ess} for \gls{fr} (\gls{SoC} model) }	&  \multirow{1}{*}{$87.55$}	&	\multirow{1}{*}{$56.91$} \\
		\hhline{---}
		\multirow{1}{*}{\rule{0pt}{1ex} 30 MW \gls{ess} for \gls{fr} (\gls{SoC} model)}	&  \multirow{1}{*}{$83.89$}	&	\multirow{1}{*}{$52.81$}	\\
		\hhline{---}
		\multirow{1}{*}{\rule{0pt}{1ex} 30 MW \gls{ess} for \gls{fr} (no \gls{SoC} model)}	&  \multirow{1}{*}{$76.98$}		&	\multirow{1}{*}{$47.41$}	\\
		\hhline{---}
		\multirow{1}{*}{\rule{0pt}{1ex} 80 MW \gls{ess} for \gls{fr} (\gls{SoC} model)}	&  \multirow{1}{*}{$80.37 $}	&	\multirow{1}{*}{$49.34$}	\\
		\hhline{===}
	\end{tabular}
	\label{table:impact_soc}
\end{table}

In order to determine the impact of communication delays in the \gls{fr} process, simulations with existing multiple delays, half of these delays, and no delays were considered. The impact is measured as a reduction in the \gls{ace}. Table \ref{table:impact_delay} presents the \gls{rmse} and \gls{mae} with respect to the ideal \gls{ace}, i.e., 0 MW. As expected, the smaller the delay, the better is the \gls{ace} performance. Note that reducing the communication delays to half their current values has approximately the same effect on the \gls{ace} as increasing 30 MW of \gls{ess} capacity for \gls{fr} with the current delay. 
\subsection{\gls{SoC} Management}\label{subsec:IVD}
To demonstrate the effect of the \gls{SoC} model in the \gls{fr} control process, the \gls{ess} capacity used for \gls{fr} was increased to 30 MW, comprising a $\pm$15 MW/30 MWh \gls{bess} and a $\pm$15 MW/3.75 MWh \gls{fess}. In the results presented in Table \ref{table:impact_delay}, it can be observed that ignoring the \gls{SoC} leads to unrealistic \gls{ace} reductions. Indeed, for the \gls{ieso} case, considering $\pm$30 MW of fast \gls{fr} without the \gls{SoC} model yields a better \gls{ace} than increasing the \gls{ess} \gls{fr} capacity to $\pm$80 MW ($\pm$40 MW/80 MWh \gls{bess}, $\pm$40 MW/10 MWh \gls{fess}) with the \gls{SoC} model, which could lead to under-procuring fast frequency response resources. 

The case of half of existing delays and 30 MW of \gls{ess} for \gls{fr}, including the \gls{SoC} model, was considered to demonstrate how a combination of reduced communication delays and increased fast \gls{fr} capacity can realistically reduce the \gls{ace}. This reduces the \gls{rmse} and \gls{mae} of \gls{ace} to 81.12 MW and 49.09 MW, respectively.
\section{Conclusions} \label{sec:conclusions}

This paper presented a validated dynamic model for long-term FR studies of a real interconnected power system including \gls{ess} facilities. The proposed estimated \gls{fr} model was designed to closely represent the frequency behaviour of a large interconnected system, and the \gls{ess} model allowed an accurate representation of the \gls{SoC} management and charging/discharging characteristics of \gls{fess} and \gls{bess}. Simulation results showed that reducing the communication delays can potentially reduce the \gls{ace} without requiring any increase in \gls{fr} capacity, and that neglecting the \gls{SoC} model of \glspl{ess} in the frequency control process yields unrealistic improvements in the \gls{ace}. 

\ifCLASSOPTIONcaptionsoff
  \newpage
\fi
\bibliographystyle{IEEEtran}
\bibliography{IEEEabrv,arxiv}
\end{document}

%% file: current_system1.pdf_tex
\begingroup%
  \makeatletter%
  \providecommand\color[2][]{%
    \errmessage{(Inkscape) Color is used for the text in Inkscape, but the package 'color.sty' is not loaded}%
    \renewcommand\color[2][]{}%
  }%
  \providecommand\transparent[1]{%
    \errmessage{(Inkscape) Transparency is used (non-zero) for the text in Inkscape, but the package 'transparent.sty' is not loaded}%
    \renewcommand\transparent[1]{}%
  }%
  \providecommand\rotatebox[2]{#2}%
  \newcommand*\fsize{\dimexpr\f@size pt\relax}%
  \newcommand*\lineheight[1]{\fontsize{\fsize}{#1\fsize}\selectfont}%
  \ifx\svgwidth\undefined%
    \setlength{\unitlength}{261.12791339bp}%
    \ifx\svgscale\undefined%
      \relax%
    \else%
      \setlength{\unitlength}{\unitlength * \real{\svgscale}}%
    \fi%
  \else%
    \setlength{\unitlength}{\svgwidth}%
  \fi%
  \global\let\svgwidth\undefined%
  \global\let\svgscale\undefined%
  \makeatother%
  \begin{picture}(1,0.81429209)%
    \lineheight{1}%
    \setlength\tabcolsep{0pt}%
    \put(0,0){\includegraphics[width=\unitlength,page=1]{current_system1.pdf}}%
  \end{picture}%
\endgroup%

%% file: filtered_ACE.pdf_tex
\begingroup%
  \makeatletter%
  \providecommand\color[2][]{%
    \errmessage{(Inkscape) Color is used for the text in Inkscape, but the package 'color.sty' is not loaded}%
    \renewcommand\color[2][]{}%
  }%
  \providecommand\transparent[1]{%
    \errmessage{(Inkscape) Transparency is used (non-zero) for the text in Inkscape, but the package 'transparent.sty' is not loaded}%
    \renewcommand\transparent[1]{}%
  }%
  \providecommand\rotatebox[2]{#2}%
  \newcommand*\fsize{\dimexpr\f@size pt\relax}%
  \newcommand*\lineheight[1]{\fontsize{\fsize}{#1\fsize}\selectfont}%
  \ifx\svgwidth\undefined%
    \setlength{\unitlength}{78.9779575bp}%
    \ifx\svgscale\undefined%
      \relax%
    \else%
      \setlength{\unitlength}{\unitlength * \real{\svgscale}}%
    \fi%
  \else%
    \setlength{\unitlength}{\svgwidth}%
  \fi%
  \global\let\svgwidth\undefined%
  \global\let\svgscale\undefined%
  \makeatother%
  \begin{picture}(1,0.15074472)%
    \lineheight{1}%
    \setlength\tabcolsep{0pt}%
    \put(0,0){\includegraphics[width=\unitlength,page=1]{filtered_ACE.pdf}}%
  \end{picture}%
\endgroup%

%% file: AGC.pdf_tex
\begingroup%
  \makeatletter%
  \providecommand\color[2][]{%
    \errmessage{(Inkscape) Color is used for the text in Inkscape, but the package 'color.sty' is not loaded}%
    \renewcommand\color[2][]{}%
  }%
  \providecommand\transparent[1]{%
    \errmessage{(Inkscape) Transparency is used (non-zero) for the text in Inkscape, but the package 'transparent.sty' is not loaded}%
    \renewcommand\transparent[1]{}%
  }%
  \providecommand\rotatebox[2]{#2}%
  \newcommand*\fsize{\dimexpr\f@size pt\relax}%
  \newcommand*\lineheight[1]{\fontsize{\fsize}{#1\fsize}\selectfont}%
  \ifx\svgwidth\undefined%
    \setlength{\unitlength}{101.48189766bp}%
    \ifx\svgscale\undefined%
      \relax%
    \else%
      \setlength{\unitlength}{\unitlength * \real{\svgscale}}%
    \fi%
  \else%
    \setlength{\unitlength}{\svgwidth}%
  \fi%
  \global\let\svgwidth\undefined%
  \global\let\svgscale\undefined%
  \makeatother%
  \begin{picture}(1,0.17651208)%
    \lineheight{1}%
    \setlength\tabcolsep{0pt}%
    \put(0,0){\includegraphics[width=\unitlength,page=1]{AGC.pdf}}%
  \end{picture}%
\endgroup%

%% file: Aggreg_TG.pdf_tex
\begingroup%
  \makeatletter%
  \providecommand\color[2][]{%
    \errmessage{(Inkscape) Color is used for the text in Inkscape, but the package 'color.sty' is not loaded}%
    \renewcommand\color[2][]{}%
  }%
  \providecommand\transparent[1]{%
    \errmessage{(Inkscape) Transparency is used (non-zero) for the text in Inkscape, but the package 'transparent.sty' is not loaded}%
    \renewcommand\transparent[1]{}%
  }%
  \providecommand\rotatebox[2]{#2}%
  \newcommand*\fsize{\dimexpr\f@size pt\relax}%
  \newcommand*\lineheight[1]{\fontsize{\fsize}{#1\fsize}\selectfont}%
  \ifx\svgwidth\undefined%
    \setlength{\unitlength}{102.23324005bp}%
    \ifx\svgscale\undefined%
      \relax%
    \else%
      \setlength{\unitlength}{\unitlength * \real{\svgscale}}%
    \fi%
  \else%
    \setlength{\unitlength}{\svgwidth}%
  \fi%
  \global\let\svgwidth\undefined%
  \global\let\svgscale\undefined%
  \makeatother%
  \begin{picture}(1,0.15804527)%
    \lineheight{1}%
    \setlength\tabcolsep{0pt}%
    \put(0,0){\includegraphics[width=\unitlength,page=1]{Aggreg_TG.pdf}}%
  \end{picture}%
\endgroup%

%% file: ESS3.pdf_tex
\begingroup%
  \makeatletter%
  \providecommand\color[2][]{%
    \errmessage{(Inkscape) Color is used for the text in Inkscape, but the package 'color.sty' is not loaded}%
    \renewcommand\color[2][]{}%
  }%
  \providecommand\transparent[1]{%
    \errmessage{(Inkscape) Transparency is used (non-zero) for the text in Inkscape, but the package 'transparent.sty' is not loaded}%
    \renewcommand\transparent[1]{}%
  }%
  \providecommand\rotatebox[2]{#2}%
  \newcommand*\fsize{\dimexpr\f@size pt\relax}%
  \newcommand*\lineheight[1]{\fontsize{\fsize}{#1\fsize}\selectfont}%
  \ifx\svgwidth\undefined%
    \setlength{\unitlength}{141.13562553bp}%
    \ifx\svgscale\undefined%
      \relax%
    \else%
      \setlength{\unitlength}{\unitlength * \real{\svgscale}}%
    \fi%
  \else%
    \setlength{\unitlength}{\svgwidth}%
  \fi%
  \global\let\svgwidth\undefined%
  \global\let\svgscale\undefined%
  \makeatother%
  \begin{picture}(1,0.88908072)%
    \lineheight{1}%
    \setlength\tabcolsep{0pt}%
    \put(0,0){\includegraphics[width=\unitlength,page=1]{ESS3.pdf}}%
  \end{picture}%
\endgroup%

%% file: abcd_index.pdf_tex
\begingroup%
  \makeatletter%
  \providecommand\color[2][]{%
    \errmessage{(Inkscape) Color is used for the text in Inkscape, but the package 'color.sty' is not loaded}%
    \renewcommand\color[2][]{}%
  }%
  \providecommand\transparent[1]{%
    \errmessage{(Inkscape) Transparency is used (non-zero) for the text in Inkscape, but the package 'transparent.sty' is not loaded}%
    \renewcommand\transparent[1]{}%
  }%
  \providecommand\rotatebox[2]{#2}%
  \newcommand*\fsize{\dimexpr\f@size pt\relax}%
  \newcommand*\lineheight[1]{\fontsize{\fsize}{#1\fsize}\selectfont}%
  \ifx\svgwidth\undefined%
    \setlength{\unitlength}{172.8698084bp}%
    \ifx\svgscale\undefined%
      \relax%
    \else%
      \setlength{\unitlength}{\unitlength * \real{\svgscale}}%
    \fi%
  \else%
    \setlength{\unitlength}{\svgwidth}%
  \fi%
  \global\let\svgwidth\undefined%
  \global\let\svgscale\undefined%
  \makeatother%
  \begin{picture}(1,0.90400839)%
    \lineheight{1}%
    \setlength\tabcolsep{0pt}%
    \put(0,0){\includegraphics[width=\unitlength,page=1]{abcd_index.pdf}}%
  \end{picture}%
\endgroup%

%% file: Max_min_bp.pdf_tex
\begingroup%
  \makeatletter%
  \providecommand\color[2][]{%
    \errmessage{(Inkscape) Color is used for the text in Inkscape, but the package 'color.sty' is not loaded}%
    \renewcommand\color[2][]{}%
  }%
  \providecommand\transparent[1]{%
    \errmessage{(Inkscape) Transparency is used (non-zero) for the text in Inkscape, but the package 'transparent.sty' is not loaded}%
    \renewcommand\transparent[1]{}%
  }%
  \providecommand\rotatebox[2]{#2}%
  \newcommand*\fsize{\dimexpr\f@size pt\relax}%
  \newcommand*\lineheight[1]{\fontsize{\fsize}{#1\fsize}\selectfont}%
  \ifx\svgwidth\undefined%
    \setlength{\unitlength}{197.78889735bp}%
    \ifx\svgscale\undefined%
      \relax%
    \else%
      \setlength{\unitlength}{\unitlength * \real{\svgscale}}%
    \fi%
  \else%
    \setlength{\unitlength}{\svgwidth}%
  \fi%
  \global\let\svgwidth\undefined%
  \global\let\svgscale\undefined%
  \makeatother%
  \begin{picture}(1,0.80808419)%
    \lineheight{1}%
    \setlength\tabcolsep{0pt}%
    \put(0,0){\includegraphics[width=\unitlength,page=1]{Max_min_bp.pdf}}%
  \end{picture}%
\endgroup%

%% file: ACE_1year1.pdf_tex
\begingroup%
  \makeatletter%
  \providecommand\color[2][]{%
    \errmessage{(Inkscape) Color is used for the text in Inkscape, but the package 'color.sty' is not loaded}%
    \renewcommand\color[2][]{}%
  }%
  \providecommand\transparent[1]{%
    \errmessage{(Inkscape) Transparency is used (non-zero) for the text in Inkscape, but the package 'transparent.sty' is not loaded}%
    \renewcommand\transparent[1]{}%
  }%
  \providecommand\rotatebox[2]{#2}%
  \newcommand*\fsize{\dimexpr\f@size pt\relax}%
  \newcommand*\lineheight[1]{\fontsize{\fsize}{#1\fsize}\selectfont}%
  \ifx\svgwidth\undefined%
    \setlength{\unitlength}{252bp}%
    \ifx\svgscale\undefined%
      \relax%
    \else%
      \setlength{\unitlength}{\unitlength * \real{\svgscale}}%
    \fi%
  \else%
    \setlength{\unitlength}{\svgwidth}%
  \fi%
  \global\let\svgwidth\undefined%
  \global\let\svgscale\undefined%
  \makeatother%
  \begin{picture}(1,0.66023393)%
    \lineheight{1}%
    \setlength\tabcolsep{0pt}%
    \put(0,0){\includegraphics[width=\unitlength,page=1]{ACE_1year1.pdf}}%
  \end{picture}%
\endgroup%

%% file: Sreg_ES_1year.pdf_tex
\begingroup%
  \makeatletter%
  \providecommand\color[2][]{%
    \errmessage{(Inkscape) Color is used for the text in Inkscape, but the package 'color.sty' is not loaded}%
    \renewcommand\color[2][]{}%
  }%
  \providecommand\transparent[1]{%
    \errmessage{(Inkscape) Transparency is used (non-zero) for the text in Inkscape, but the package 'transparent.sty' is not loaded}%
    \renewcommand\transparent[1]{}%
  }%
  \providecommand\rotatebox[2]{#2}%
  \newcommand*\fsize{\dimexpr\f@size pt\relax}%
  \newcommand*\lineheight[1]{\fontsize{\fsize}{#1\fsize}\selectfont}%
  \ifx\svgwidth\undefined%
    \setlength{\unitlength}{252bp}%
    \ifx\svgscale\undefined%
      \relax%
    \else%
      \setlength{\unitlength}{\unitlength * \real{\svgscale}}%
    \fi%
  \else%
    \setlength{\unitlength}{\svgwidth}%
  \fi%
  \global\let\svgwidth\undefined%
  \global\let\svgscale\undefined%
  \makeatother%
  \begin{picture}(1,0.66094751)%
    \lineheight{1}%
    \setlength\tabcolsep{0pt}%
    \put(0,0){\includegraphics[width=\unitlength,page=1]{Sreg_ES_1year.pdf}}%
  \end{picture}%
\endgroup%

%% file: SREG_1year.pdf_tex
\begingroup%
  \makeatletter%
  \providecommand\color[2][]{%
    \errmessage{(Inkscape) Color is used for the text in Inkscape, but the package 'color.sty' is not loaded}%
    \renewcommand\color[2][]{}%
  }%
  \providecommand\transparent[1]{%
    \errmessage{(Inkscape) Transparency is used (non-zero) for the text in Inkscape, but the package 'transparent.sty' is not loaded}%
    \renewcommand\transparent[1]{}%
  }%
  \providecommand\rotatebox[2]{#2}%
  \newcommand*\fsize{\dimexpr\f@size pt\relax}%
  \newcommand*\lineheight[1]{\fontsize{\fsize}{#1\fsize}\selectfont}%
  \ifx\svgwidth\undefined%
    \setlength{\unitlength}{252bp}%
    \ifx\svgscale\undefined%
      \relax%
    \else%
      \setlength{\unitlength}{\unitlength * \real{\svgscale}}%
    \fi%
  \else%
    \setlength{\unitlength}{\svgwidth}%
  \fi%
  \global\let\svgwidth\undefined%
  \global\let\svgscale\undefined%
  \makeatother%
  \begin{picture}(1,0.66023393)%
    \lineheight{1}%
    \setlength\tabcolsep{0pt}%
    \put(0,0){\includegraphics[width=\unitlength,page=1]{SREG_1year.pdf}}%
  \end{picture}%
\endgroup%

%% file: OutTG_1year.pdf_tex
\begingroup%
  \makeatletter%
  \providecommand\color[2][]{%
    \errmessage{(Inkscape) Color is used for the text in Inkscape, but the package 'color.sty' is not loaded}%
    \renewcommand\color[2][]{}%
  }%
  \providecommand\transparent[1]{%
    \errmessage{(Inkscape) Transparency is used (non-zero) for the text in Inkscape, but the package 'transparent.sty' is not loaded}%
    \renewcommand\transparent[1]{}%
  }%
  \providecommand\rotatebox[2]{#2}%
  \newcommand*\fsize{\dimexpr\f@size pt\relax}%
  \newcommand*\lineheight[1]{\fontsize{\fsize}{#1\fsize}\selectfont}%
  \ifx\svgwidth\undefined%
    \setlength{\unitlength}{252bp}%
    \ifx\svgscale\undefined%
      \relax%
    \else%
      \setlength{\unitlength}{\unitlength * \real{\svgscale}}%
    \fi%
  \else%
    \setlength{\unitlength}{\svgwidth}%
  \fi%
  \global\let\svgwidth\undefined%
  \global\let\svgscale\undefined%
  \makeatother%
  \begin{picture}(1,0.66023393)%
    \lineheight{1}%
    \setlength\tabcolsep{0pt}%
    \put(0,0){\includegraphics[width=\unitlength,page=1]{OutTG_1year.pdf}}%
  \end{picture}%
\endgroup%

%% file: FESS_FASTSP.pdf_tex
\begingroup%
  \makeatletter%
  \providecommand\color[2][]{%
    \errmessage{(Inkscape) Color is used for the text in Inkscape, but the package 'color.sty' is not loaded}%
    \renewcommand\color[2][]{}%
  }%
  \providecommand\transparent[1]{%
    \errmessage{(Inkscape) Transparency is used (non-zero) for the text in Inkscape, but the package 'transparent.sty' is not loaded}%
    \renewcommand\transparent[1]{}%
  }%
  \providecommand\rotatebox[2]{#2}%
  \newcommand*\fsize{\dimexpr\f@size pt\relax}%
  \newcommand*\lineheight[1]{\fontsize{\fsize}{#1\fsize}\selectfont}%
  \ifx\svgwidth\undefined%
    \setlength{\unitlength}{419.49375627bp}%
    \ifx\svgscale\undefined%
      \relax%
    \else%
      \setlength{\unitlength}{\unitlength * \real{\svgscale}}%
    \fi%
  \else%
    \setlength{\unitlength}{\svgwidth}%
  \fi%
  \global\let\svgwidth\undefined%
  \global\let\svgscale\undefined%
  \makeatother%
  \begin{picture}(1,1.29052928)%
    \lineheight{1}%
    \setlength\tabcolsep{0pt}%
    \put(0,0){\includegraphics[width=\unitlength,page=1]{FESS_FASTSP.pdf}}%
  \end{picture}%
\endgroup%

%% file: BESS_FASTSP.pdf_tex
\begingroup%
  \makeatletter%
  \providecommand\color[2][]{%
    \errmessage{(Inkscape) Color is used for the text in Inkscape, but the package 'color.sty' is not loaded}%
    \renewcommand\color[2][]{}%
  }%
  \providecommand\transparent[1]{%
    \errmessage{(Inkscape) Transparency is used (non-zero) for the text in Inkscape, but the package 'transparent.sty' is not loaded}%
    \renewcommand\transparent[1]{}%
  }%
  \providecommand\rotatebox[2]{#2}%
  \newcommand*\fsize{\dimexpr\f@size pt\relax}%
  \newcommand*\lineheight[1]{\fontsize{\fsize}{#1\fsize}\selectfont}%
  \ifx\svgwidth\undefined%
    \setlength{\unitlength}{417.60299952bp}%
    \ifx\svgscale\undefined%
      \relax%
    \else%
      \setlength{\unitlength}{\unitlength * \real{\svgscale}}%
    \fi%
  \else%
    \setlength{\unitlength}{\svgwidth}%
  \fi%
  \global\let\svgwidth\undefined%
  \global\let\svgscale\undefined%
  \makeatother%
  \begin{picture}(1,1.31093404)%
    \lineheight{1}%
    \setlength\tabcolsep{0pt}%
    \put(0,0){\includegraphics[width=\unitlength,page=1]{BESS_FASTSP.pdf}}%
  \end{picture}%
\endgroup%